\documentclass[%
pre,reprint,
 amsmath,amssymb,
 aps,
]{revtex4-1}

\usepackage{graphicx}
\usepackage{dcolumn}
\usepackage{bm}
\usepackage{textcomp}
\usepackage[pdftex,colorlinks=true,bookmarks=false,citecolor=blue,urlcolor=blue]{hyperref}
\usepackage{cancel}
\usepackage{color}
\newcolumntype{?}{!{\vrule width 1.1pt}}

\allowdisplaybreaks
\newcommand{\mrm}[1]{\mathrm{#1}}

\newcommand{\rmi}{\mathrm{i}} 

\newcommand{\Evec}{\mathbf{E}}	
\newcommand{\Bvec}{\mathbf{B}}	
\newcommand{\Jvec}{\mathbf{J}}	

\newcommand{\rvec}{\mathbf{r}}
\newcommand{\evec}{\mathbf{e}}	

\DeclareSymbolFont{lettersA}{U}{txmia}{m}{it}
\DeclareMathSymbol{\real}{\mathord}{lettersA}{"92} 
\DeclareMathSymbol{\cplx}{\mathord}{lettersA}{"83} 

\newcommand{\overbar}[1]{\mkern 1.5mu\overline{\mkern-1.5mu#1\mkern-1.5mu}\mkern 1.5mu}

\begin{document}


\title{Broadband terahertz emission from two-color fs-laser-induced microplasmas}\

\author{I. Thiele$^1$}
\email{illia-thiele@web.de}
\author{P. Gonz{\'a}lez de Alaiza Mart{\'i}nez$^1$}
\author{R. Nuter$^1$}
\author{A. Nguyen$^2$}
\author{L. Berg\'e$^2$}
\author{S. Skupin$^{1,3}$}
\address{$^1$Univ.~Bordeaux - CNRS - CEA, Centre Lasers Intenses et Applications, UMR 5107, 33405 Talence, France\\
$^2$CEA/DAM {\^I}le-de-France, Bruy\`eres-le-Ch\^atel, 91297 Arpajon, France\\
$^3$Institut Lumi\`ere Mati\`ere, UMR 5306 Universit\'e Lyon 1 - CNRS, Universit\'e de Lyon, 69622 Villeurbanne, France}


\date{\today}

\begin{abstract}
We investigate terahertz emission from two-color fs-laser-induced microplasmas. Under strongest focusing conditions, microplasmas are shown to act as point-sources for broadband terahertz-to-far-infrared radiation, where the emission bandwidth is determined by the plasma density. Semi-analytical modeling allows us to identify scaling laws with respect to important laser parameters. In particular, we find that the optical-to-THz conversion efficiency crucially depends on the focusing conditions. We use this insight to demonstrate by means of Maxwell-consistent 3D simulations, that for only 10-$\mu$J laser energy a conversion efficiency well above $10^{-4}$ can be achieved. 
\end{abstract}

\maketitle


Terahertz (THz) sources are essential for various applications such as imaging and time-domain spectroscopy or control of matter by THz waves \cite{Tonouchi, 0034-4885-70-8-R02, Kampfrath}. A promising approach to generate broadband THz radiation is to employ laser-induced gas plasmas~\cite{PhysRevLett.98.235002,Kim}. Spectral properties and the absence of irreversible material damage make them interesting alternatives to conventional THz sources such as photo-conductive switches or quantum cascade lasers. In order to miniaturize such THz sources, it has been proposed to exploit laser-induced microplasmas~\cite{Buccheri:15} that allow to use smaller driving lasers. Such microplasmas are created when a fs-laser pulse is focused tightly, down to the diffraction limit, into a gas. The strong focusing leads to laser intensities above $10^{14}$~W/cm$^2$ already for $\mu$J pulses, which is sufficient to singly ionize the neutral gas, and to produce electron densities above $10^{19}$~cm$^{-3}$ at ambient pressure. Thus, gas-based THz generation mechanisms are accessible with such small pulse energies.

\begin{figure}[b]
	\centering
	\includegraphics[width=.7\columnwidth]{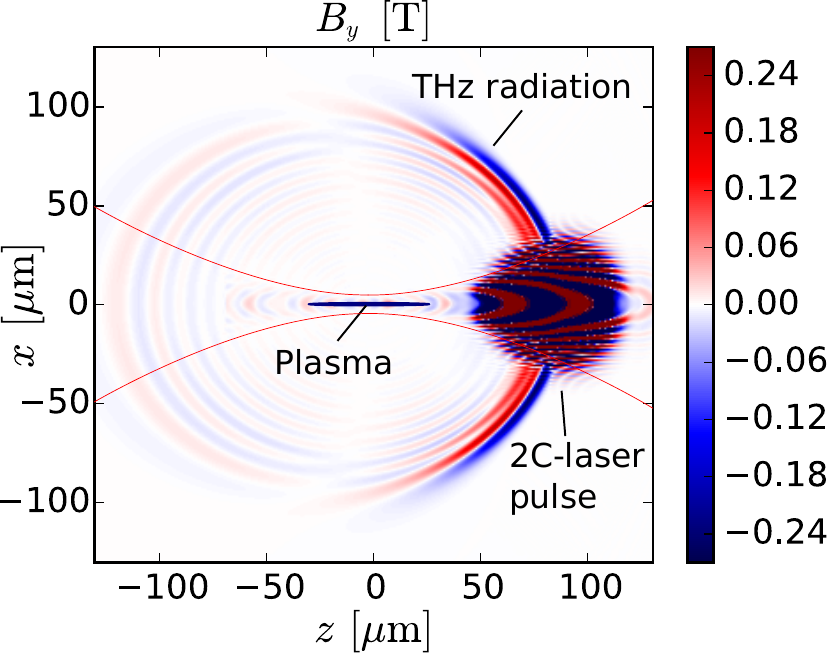}
	\caption{Schematic illustration of two-color fs-laser-induced THz generation in a gas: A snapshot of the magnetic field $B_y$ after the interaction of the laser pulse with the laser-induced plasma is presented. The two-color pulse and the generated THz radiation are indicated in the graph. The waist of the focused laser is tracked by the dark red lines, the position of the generated plasma is marked as a blue oval.}
	\label{fig:scheeme}
\end{figure}

Recently, THz radiation from single-color fs-laser-induced microplasmas has been demonstrated~\cite{Buccheri:15}. In this case, THz emission is produced by longitudinal low-frequency currents driven by the ponderomotive force, usually referred to as transition Cherenkov (TC) mechanism~\cite{PhysRevLett.98.235002}. However, the laser-to-THz conversion efficiency for this scenario has been shown to saturate, even for increasing pulse energies, below $10^{-6}$~\cite{PhysRevE.94.063202}. 
Opaqueness of the plasma for THz frequencies has been shown to play the key role for the saturation behavior. The low efficiency results from the low excitation strength by ponderomotive forces for non-relativistic intensities. 
To overcome this limitations, we investigate the potential of the ionization current (IC) mechanism in microplasmas, where THz emission is driven by two-color laser pulses~\cite{Kim}~(see Fig.~\ref{fig:scheeme}). 
Effective THz emission by the IC mechanism requires a temporal asymmetry in the driving laser field that can be achieved in a straightforward manner by admixing the second harmonic (SH) to the fundamental harmonic (FH) frequency of the laser pulse.  
Then, the electric force acting on the electrons excites THz-radiating ionization currents parallel to the laser electric field, i.e. mainly transverse to the laser propagation direction. As we show in this article, it is this transverse nature of the ionization currents what is advantageous for the up-scaling of the laser-to-THz conversion efficiency. Moreover, THz emission from two-color fs-laser-induced microplasmas turns out to be extremely broadband, covering the spectral range from below 1~THz up to the plasma frequency, which is about 50~THz for a singly ionized gas at ambient pressure. 

\begin{figure*}
	\includegraphics[width=0.8\textwidth]{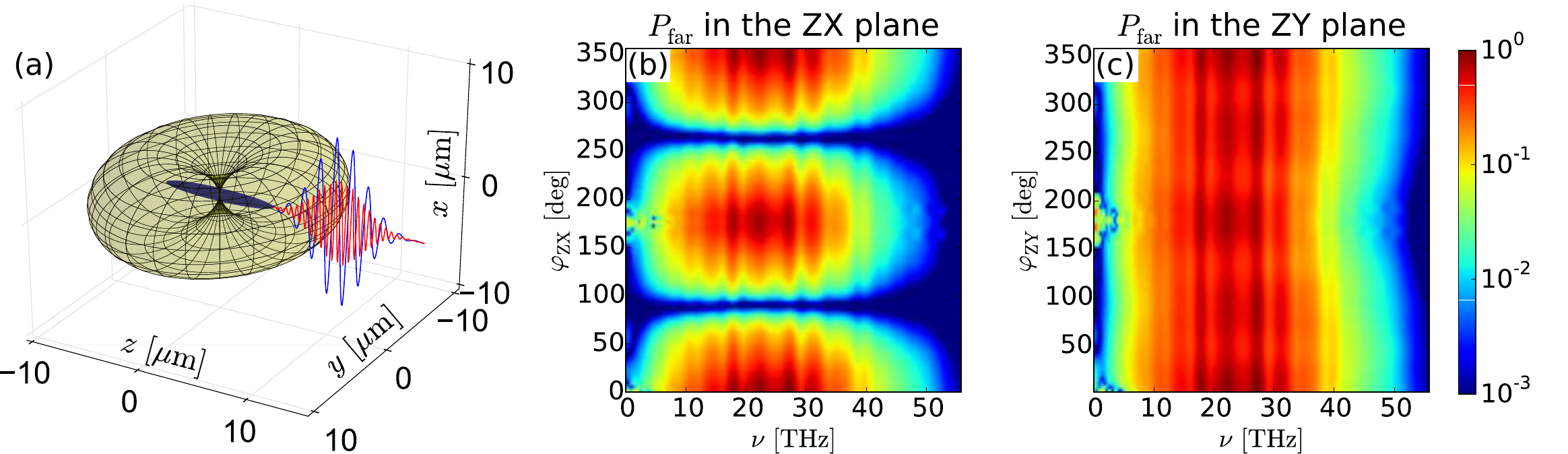}
	\caption{Two-color 0.17-$\mu$J Gaussian laser pulse (20\% energy in the SH, $t_0=50$~fs) focused down to $w_0=\lambda_\mathrm{FH}=0.8~\mu$m in argon with initial neutral density $n_\mrm{a}=3\cdot10^{19}$~cm$^{-3}$ ($p\approx1$~bar). The radiation pattern for THz frequencies (assuming a point source at $\rvec=0$) is sketched in (a) (green surface) as well as the plasma volume (blue surface). Both FH (blue line) and SH (red line) electric fields are $x$-polarized. In (b-c) the simulated THz radiation profiles in terms of the far field power spectra $P_{\rm far}$ in ZX ($y=0$) and ZY ($x=0$) planes are presented. We count the azimuthal angle $\varphi_\mathrm{ZX}$ and $\varphi_\mathrm{ZY}$ counterclockwise from the $z$-axis respectively. The laser pulse is propagating in positive $z$ direction. $P_{\rm far}$ in the ZX ($y=0$) and YX ($z=0$, not shown) are identical due to the cylindrical symmetry of the emission profile with respect to the $x$-axis.}
	\label{fig:radprof_pol}
\end{figure*}

The article is organized as follows: After summarizing the numerical model, we first analyze sub-$\mu$J two-color pulses that produce smallest possible microplasmas. Resulting emission profiles are shown to provide a unique feature of point-like THz emission giving the opportunity to observe the polarization of the THz generating current directly from the THz emission profile. Broadband THz spectra are shown to be strongly dependent on the gas pressure. Secondly, we identify important scaling laws of the IC mechanism in microplasmas by means of a simplified model. In particular, optimization of the laser-to-THz conversion efficiency with respect to the pump pulse energy is addressed. Finally, we confront our predictions with rigorous Maxwell-consistent 3D simulations, and we demonstrate conversion efficiencies above $10^{-4}$ for about 5-$\mu$J laser pulse energy.


Throughout this article, we consider argon gas and fundamental laser wavelength of 800~nm.
Under strongest focusing conditions, a one to several micrometer thick and tens of micrometer long microplasma with peak electron densities above $10^{19}$~cm$^{-3}$ can be created~\cite{PhysRevE.94.063202}. The tight focusing as well as the opaqueness of such plasma for THz frequencies requires a Maxwell-consistent modeling beyond the often employed unidirectional approach~\cite{Kolesik:pre:70:036604,PhysRevLett.105.053903}. 
To this end, we couple Maxwell's equations to a Drude model for the macroscopic current density $\Jvec$ 
\begin{align}
	\partial_t \Jvec + \nu_\mrm{ei}\Jvec &= \frac{q_\mrm{e}^2}{m_\mrm{e}} n_\mrm{e} \Evec\label{eq:current}\,\mbox{,}
\end{align}
where $q_\mrm{e}$ is the charge and $m_\mrm{e}$ the mass of an electron. The electron density $n_\mrm{e}$ is computed by means of rate equations employing the tunnel ionization rate \cite{Ammosov-1986-Tunnel,PhysRevA.64.013409}. The electron-ion collision frequency $\nu_\mrm{ei}$ depends on the ion densities and the electron energy density~\cite{Huba2013}. Our model appears as the lowest order of a multiple-scale expansion developed in \cite{PhysRevE.94.063202}, and fully comprises the IC mechanism. In App.~\ref{app:num}, we briefly review this model that is solved by our code ARCTIC and benchmark it against more rigorous particle-in-cell simulations. 

The driving laser pulse is defined by its transverse vacuum electric field at focus according to
\begin{equation}
\begin{split}
	&\Evec_{\mrm{L},\perp}(\rvec_\perp, z=0, t) = \exp\!\left(-\frac{|\rvec_\perp|^2}{w_0^2}-\frac{t^2}{t_0^2}\right)\\
	&\quad\times\bigg[E_{\omega}\cos\!\left(\omega_\mrm{L}t\right)\evec_\mrm{FH}+E_{2\omega}\cos\!\left(2\omega_\mrm{L}t+\phi\right)\evec_\mrm{SH}\bigg]\mbox{,}
\end{split}
\end{equation}
where $\rvec_\perp=(x,y)^\mrm{T}$ and $z$ account for the transverse and longitudinal spatial coordinates, resp., $t$ is the time coordinate, $w_0$ the vacuum focal beam width, $t_0$ the pulse duration, $E_{\omega}$,  $E_{2\omega}$ the FH or SH electric field amplitudes, resp., $\omega_\mrm{L}$ the FH laser frequency, $\phi$ the SH relative phase angle, and the unit vectors $\evec_\mrm{FH}$, $\evec_\mrm{SH}$ define the (linear) polarization direction of the FH or SH, respectively. The laser pulse is propagating in the positive $z$ direction, and the origin of the coordinate system is chosen at its vacuum focal point. By defining the tightly focused laser pulse via its properties at the vacuum focus we follow the algorithm described in~\cite{Thiele20161110}.

Let us start with the smallest microplasma, when a two-color 50-fs pump pulse is focused down to the diffraction limit (here $w_0 = 0.8\,\mu$m). To estimate the optimum laser pulse parameters for THz generation it is common to consider the excitation source term $n_\mrm{e}\Evec_\mrm{L}$ (see App.~\ref{app:phase}-\ref{app:amplitude} for details): The optimum phase angle is known to be  $\phi=\pi/2$~\cite{Kim}. Moreover, one can expect that close to full single ionization a ratio of $1/2$ between SH and FH field amplitudes gives the highest THz yield~\cite{martinez:prl:114:183901}, as well as parallel polarization $\evec_\mrm{FH}\parallel\evec_\mrm{SH}$~\cite{olga_pol}. Thus, we fix $E_{\omega}=40$~GV/m, $E_{\omega}=20$~GV/m, and $\evec_\mrm{FH}=\evec_\mrm{SH}=\evec_x$, which is just sufficient to fully singly ionize the argon gas at ambient pressure. Then, the pulse energy is only 0.17~$\mu$J, and the laser-induced plasma is approximately 8~$\mu$m long. For such short interaction length, the laser propagation is practically unaffected and follows almost the vacuum case, where the phase angle $\phi$ is evolving along $z$ due to the particular frequency dependence of the Gouy phase. The optimum value of $\phi=\pi/2$ is reached only in the focal plane, and the longitudinal variation of $\phi$ further reduces the effective length of the THz source to approximately $4\,\mu$m. Therefore, the source is small compared to THz wavelengths and the microplasma acts as a point source, i.e., a dipole. 
Because FH and SH are $x$-polarized, the dipole-like THz radiation profile forms a torus with a hole along the $x$ axis as sketched in Fig.~\ref{fig:radprof_pol}(a). Simulations confirm the point-like emission as presented by the angularly resolved THz far-field spectra for two different planes in Fig.~\ref{fig:radprof_pol}(b-c). We regain the characteristic hole of the torus in (b) at $\varphi_\mrm{ZX}=90^\circ,270^\circ$. Cutting the torus along the ZY plane, like a bagel, we find as expected the $\varphi_\mrm{ZY}$-independent radiation profile in (c). The characteristic THz radiation pattern from smallest microplasmas could allows a direct determination of the THz-driving-current polarization in future experiments. 

\begin{figure}[b]
   \includegraphics[width=1.0\columnwidth]{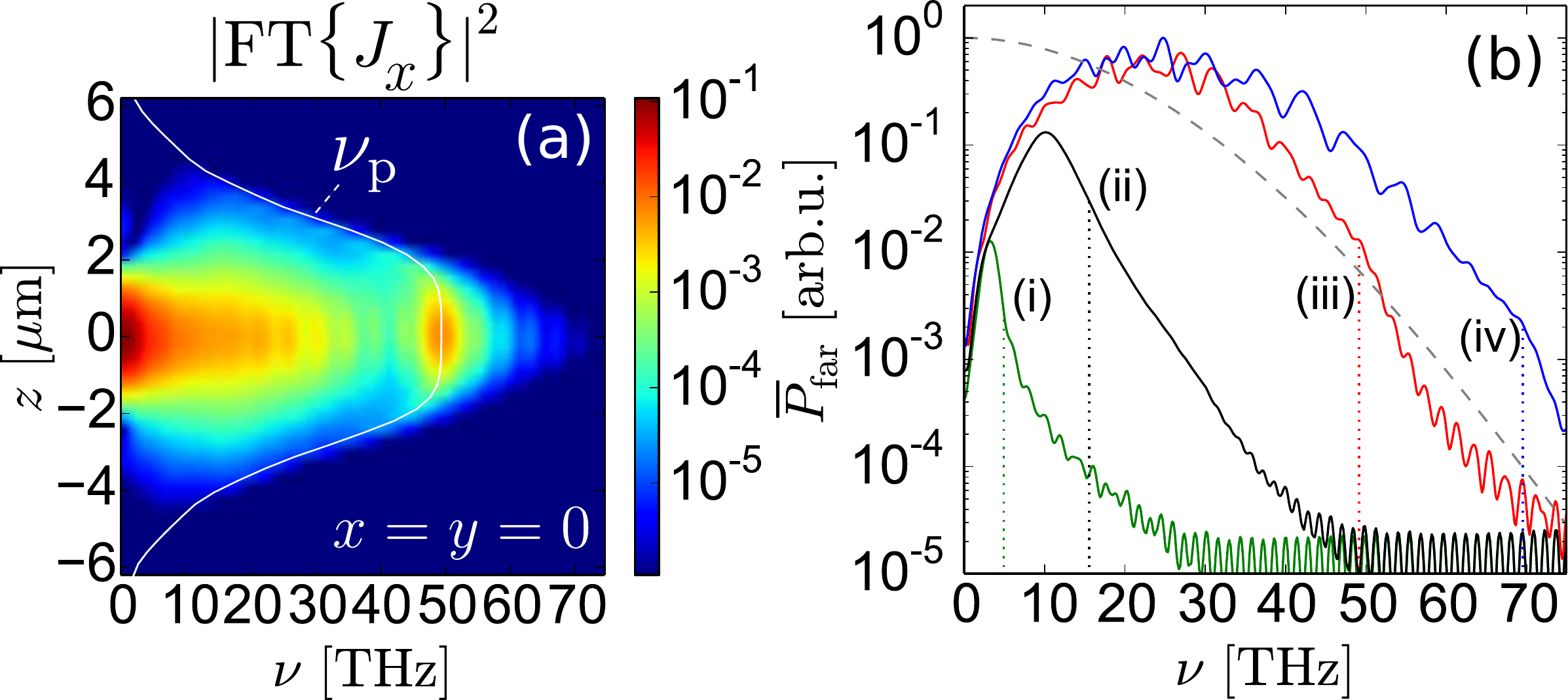}
   \caption{(a) Power spectrum of the THz emitting current $J_x$ on the optical axis (a) normalized to its respective value at $\omega_\mrm{L}$. The laser and gas parameters are the same as in Fig.~\ref{fig:radprof_pol}. The local plasma frequency $\nu_\mrm{p}$ (after ionization) is indicated by the white line. (b) The angularly integrated far-field power spectrum $\overbar{P}_\mrm{far}$ in THz spectral range for $p\approx10^{-2}$~bar~(i, green line), $p\approx10^{-1}$~bar~(ii, black line), $p\approx1$~bar~(iii, red line) and $p\approx2$~bar~(vi, blue line). The maximum plasma frequency (at focus) is marked by dotted lines for each curve. Additionally, the normalized power spectrum of the excitation $n_\mrm{e}E_{\mrm{L},x}$ at focus is presented as light gray dashed line.}
   \label{fig:spec_broad}
\end{figure}

As shown in Fig.~\ref{fig:radprof_pol}(b-c), broadband THz-to-far-infrared emission up to $\omega/(2\pi)=\nu\sim50$~THz is observed from two-color-laser-induced microplasmas. Such a broad emission spectrum requires a broadband THz emitting current. The power spectrum of the THz current on the optical axis is presented in Fig.~\ref{fig:spec_broad}(a). At focus ($z=0$), the plasma is mostly excited up to about 50~THz, the local plasma frequency $\nu_\mrm{p}=\sqrt{q_\mrm{e}^2 n_\mrm{e}/(m_\mrm{e} \epsilon_0)}/(2\pi)$, with vacuum permittivity $\epsilon_0$. The white line in Fig.~\ref{fig:spec_broad}(a) specifies $\nu_\mrm{p}$ along the $z$-axis, as it evolves with the plasma density. It appears that the current spectrum is locally broadened up to $\nu_\mrm{p}$. The same observation holds in the transverse directions (not shown). This broadening of the current power spectrum affects the THz emission properties: Figure~\ref{fig:spec_broad}(b) shows the angularly integrated THz far-field spectrum for different gas pressures, i.e., different maximum plasma densities. According to the gas pressure, the spectrum broadens up to the maximum plasma frequency $\nu_\mrm{p}$, as indicated by the vertical dotted lines. Similar broadening with increasing gas pressure has been already observed experimentally for much longer plasmas \cite{PhysRevLett.105.053903}. Here, the broadening was explained by laser propagation effects. Our Maxwell-consistent simulations show, however, that such laser propagation effects can be ruled out for smallest microplasmas. Thus, the present broadening originates rather from the particular configuration involving sharp spatial gradients in the electron density.

The excitation of the plasma by the strongly focused two-color fs-laser pulse as considered so far is quite effective: The THz current power spectrum in Fig.~\ref{fig:spec_broad}(a) is just one order of magnitude below the current at the laser frequency $\omega_\mrm{L}$. However, the laser-to-THz conversion efficiency $\eta_\mrm{THz}$, defined as the energy emitted at all angles in the range from 0 to 60~THz divided by the pump pulse energy, is of the order of $10^{-6}$. 
This is already one order of magnitude larger than what can be reached with single-color laser pulses with same energy and exploiting the TC mechanism. While single-color laser pulses do not provide a scaling of $\eta_\mrm{THz}$ beyond $10^{-6}$~\cite{PhysRevE.94.063202}, it is worth investigating the upscaling for multi-$\mu$J two-color pulses.

\begin{figure}[b]
	\centering
	\includegraphics[width=.8\columnwidth]{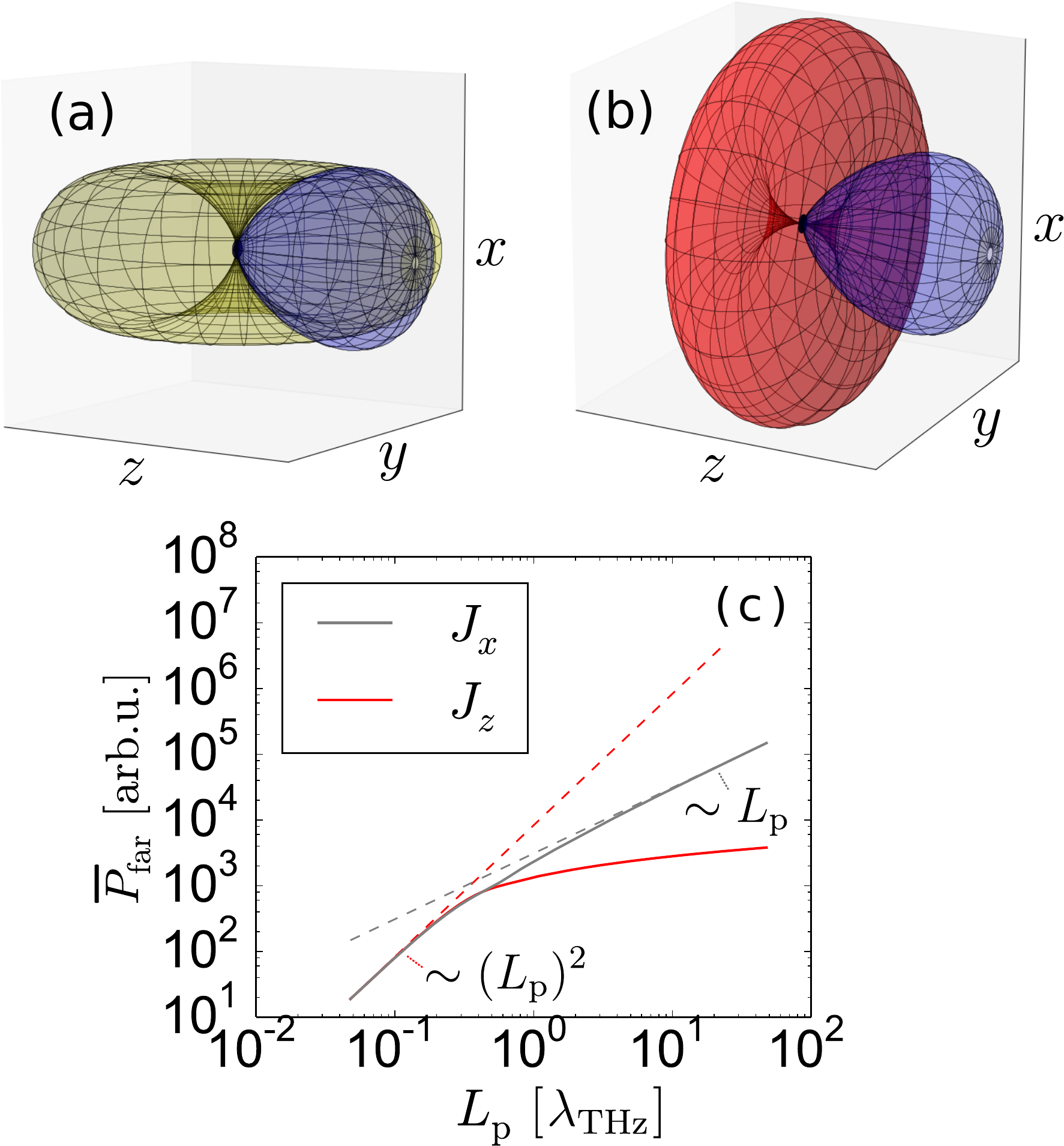}
	\caption{Visualization of the angular dependency of the factors $F_{\Jvec_0}$ and $F_{D_\mrm{p}}$ in Eq.~(\ref{eq:Pfar_approx}). The thin plasma is oriented along the $z$-axis. In (a), a transverse current $\Jvec_0\parallel\evec_x$ is considered, and the toroidal $F_{\Jvec_0}$ has its hole oriented along the $x$-direction (green surface). In (b), a longitudinal current $\Jvec_0\parallel\evec_z$ is considered, and the toroidal $F_{\Jvec_0}$ has its hole oriented along the $z$-direction (red surface). The plasma length dependent $F_{L_\mrm{p}}$ is illustrated in (a,b) in blue for $L_\mrm{p}=1.6\lambda_\mrm{THz}$. Its fig-shaped lobe is oriented in the laser propagation direction along $z$. The product between $F_{L_\mrm{p}}$ and $F_{\Jvec_0}$ results in the emission profile. In (c), THz emission scaling in terms of the angularly integrated far-field power spectrum $\overbar{P}_\mrm{far}$  depending on $L_\mrm{p}$ for a transverse ($J_x$) and longitudinal current ($J_z$) is shown. The scaling for short plasmas (dashed dark red line, $J_x$ and $J_z$) and long plasmas (dashed light gray line, $J_x$ only) is presented.}
	\label{fig:rad_prof_model_1}
\end{figure}

To this end, we resort to a simple model that is derived in more detail in App.~\ref{app:cyl},~{\color{red}G}. In short, we assume a THz radiating current in the plasma volume with length $L_\mrm{p}$ and width $D_\mrm{p}$ which is translational invariant in the co-moving pulse frame of the laser: $\Jvec(\rvec_\perp,z,t)=\Jvec_0(t-z/c)$, where $c$ is the speed of light, $|z|<L_\mrm{p}/2$, and $|\rvec_\perp|<D_\mrm{p}/2$. 
When the width of the current profile $D_\mrm{p}$ is small compared to the emission wavelength $\lambda_\mrm{THz}=c/\nu_\mrm{THz}$, it can be shown that the far-field power spectrum $P_\mrm{far}$ separates into three factors,
\begin{equation}
	P_\mrm{far} \approx \underbrace{\frac{\mu_0\omega^2 D_\mrm{p}^4}{256cr^2}}_{=:F_{D_\mrm{p}}(r)}\underbrace{\left|\evec_r\times\hat{\Jvec}_0(\omega)\right|^2}_{=:F_{\Jvec_0}(\theta,\phi)} \underbrace{\left|\int\limits_{-\frac{L_\mrm{p}}{2}}^{\frac{L_\mrm{p}}{2}}e^{\rmi\frac{\omega}{c}\left(1-\cos\theta\right)z^\prime}\,dz^\prime\right|^2}_{=:F_{L_\mrm{p}}(\theta)}\,\mbox{,} \label{eq:Pfar_approx}
\end{equation}
where we switched to standard (ISO) spherical coordinates $(r,\theta,\phi)$, and $\evec_r$ is the corresponding unit vector in the direction of increasing $r$. The first factor $F_{D_\mrm{p}}$ contains the width of the plasma, and has no angular dependency. The second factor $F_{\Jvec_0}$ contains information about the current, in particular its orientation, and the third factor $F_{L_\mrm{p}}$ contains the dependency on the length of the plasma. 
Both $F_{\Jvec_0}$ and $F_{L_\mrm{p}}$ depend on the detection angle, which is visualized in Fig.~\ref{fig:rad_prof_model_1}.
For a fixed orientation of the current, $F_{\Jvec_0}$ has the toroidal shape as expected for the emission profile of a point source. 
For a transverse current as in the previous examples, i.e., $\Jvec_0\parallel\evec_x$, the hole of the torus is oriented along the $x$-direction, as shown by the green surface in Fig.~\ref{fig:rad_prof_model_1}(a). For a point source, i.e., $L_\mrm{p}\ll\lambda_\mrm{THz}$, the factor $F_\mrm{L_\mrm{p}}$ is constant, $F_\mrm{L_\mrm{p}} \propto L_\mrm{p}^2$, and gives just a sphere in our present visualization (not shown). For larger $L_\mrm{p}$, i.e., when passing to a line source, $F_{L_\mrm{p}}$ becomes a fig-shaped structure oriented in the laser propagation direction as visualized by the blue surface in Fig.~\ref{fig:rad_prof_model_1}~(a). The radiated flux is then proportional to the product of $F_{\Jvec_0}$ and $F_\mrm{L_\mrm{p}}$. Obviously, there is some good angular overlap between the two quantities in the case of a transverse current.
The solid light gray line in Fig.~\ref{fig:rad_prof_model_1}(c) shows the predicted dependency of the angularly integrated far-field power spectrum $\overbar{P}_\mrm{far}$ on $L_\mrm{p}$. For $L_\mrm{p}\ll\lambda_\mrm{THz}$, we find $\overbar{P}_\mrm{far} \propto L_\mrm{p}^2$, whereas for  $L_\mrm{p}\gg\lambda_\mrm{THz}$ and transverse current one gets $\overbar{P}_\mrm{far} \propto L_\mrm{p}$.

It is, of course, also possible to evaluate this simple model in Eq.~(\ref{eq:Pfar_approx}) for a longitudinal current $\Jvec_0\parallel\evec_z$, as it occurs in the TC mechanism~\cite{PhysRevLett.98.235002,Buccheri:15,PhysRevE.94.063202}. In this case, $F_\mrm{L_\mrm{p}}$ remains unchanged, but the toroidal $F_{\Jvec_0}$ is oriented differently, as shown in 
Fig.~\ref{fig:rad_prof_model_1}(b). Since the overlap between $F_{\Jvec_0}$ and $F_\mrm{L_\mrm{p}}$ is crucial for efficient THz emission, 
it is a priori clear that relying on a longitudinal current is detrimental when passing to longer plasmas. indeed, the corresponding
angularly integrated far-field power spectrum $\overbar{P}_\mrm{far}$ shown in Fig.~\ref{fig:rad_prof_model_1}(c) (solid dark red line) confirms a sublinear 
scaling with $L_\mrm{p}$ for a longitudinal current.

\begin{figure}[t]
\includegraphics[width=\columnwidth]{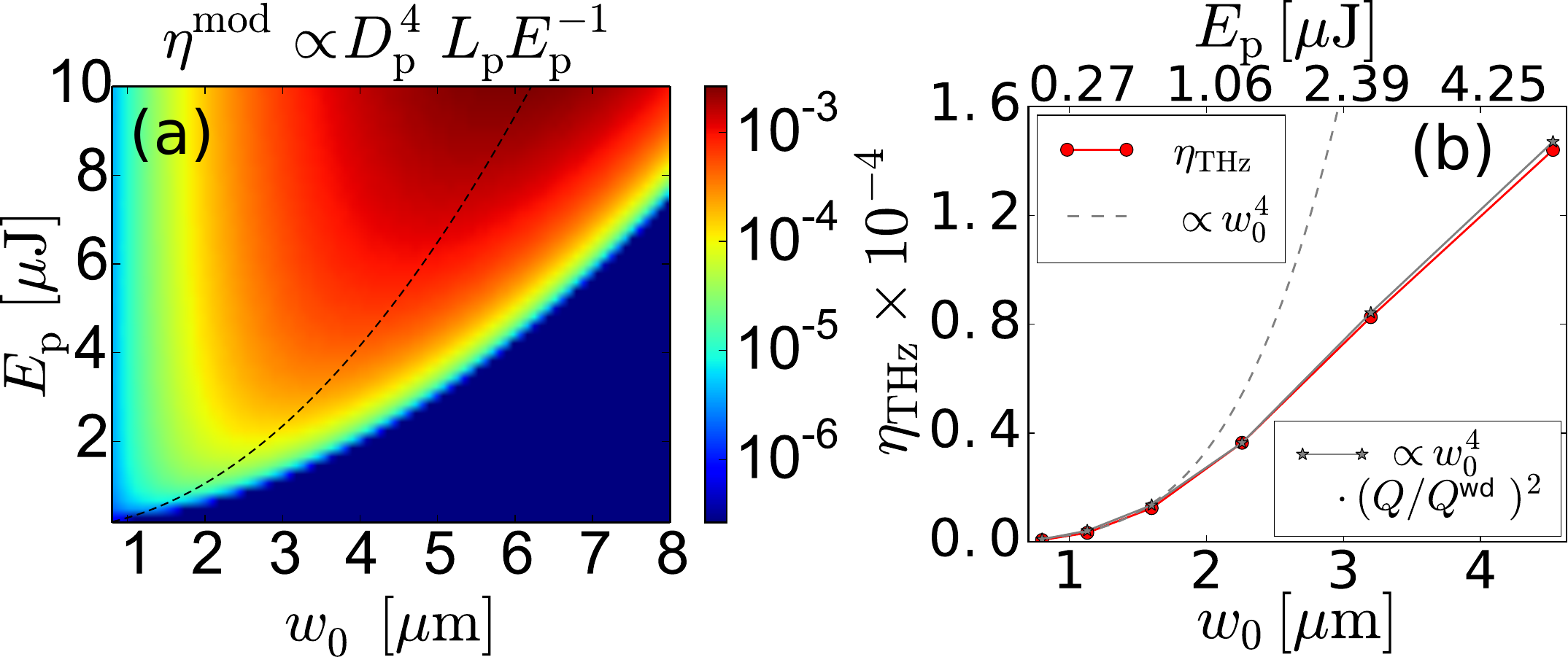}
\caption{(a) Estimation of the laser-to-THz conversion efficiency based on the simple model Eq.~(\ref{eq:Pfar_approx}) versus pulse energy $E_p$ and beam width at focus $w_0$. The black dashed line specifies $E_\mrm{p}=0.17\,\mu\mrm{J}\times(w_0/0.8\,\mu\mrm{m})^2$, 
thus it scales the laser pulse of Fig.~\ref{fig:radprof_pol} to larger 
focal beam widths $w_0$ while keeping the peak intensity constant. As the figure shows, the optimum beam width $w_0$ for a given laser pulse energy is very close to this line.
(b) Scaling of laser-to-THz conversion efficiency $\eta_\mrm{THz}$ following the black dashed line in (a) obtained from rigorous numerical simulations (solid dark red line), model disregarding laser defocusing (light dashed gray line) and model corrected for laser defocusing (solid light gray line, see text).}
\label{fig:rad_prof}
\end{figure}

These intermediate results obtained from our simple model Eq.~(\ref{eq:Pfar_approx}) suggest that for the IC mechanism, where THz radiating currents are transverse, increasing the laser-to-THz conversion efficiency $\eta_\mrm{THz}$ by passing to larger plasma volumes should be possible. 
As long as the considered plasmas are thin ($D_{\mrm{p}} \ll \lambda_{\mrm{THz}}$), emitters at different transverse positions radiate coherently.
Thus, the emitted THz pulse energy should scale with the emitting plasma surface squared, i.e. $\propto D_\mrm{p}^4$. For a line source, according to our model we can therefore expect that $\eta_\mrm{THz} \propto D_\mrm{p}^4 L_\mrm{p} / E_p$, where $E_p$ is the laser pulse energy.
We can use this rough estimate for the conversion efficency to optimize pulse energy $E_p$ versus beam width at focus $w_0$.
Results are shown in Fig.~\ref{fig:rad_prof}(a), where 
$D_\mrm{p}$ and $L_\mrm{p}$ have been determined as the plasma width and plasma length defined by the value of electron density that is 10~\% of the initial atom density. The 
computation of the electron density profile has been performed employing 
the paraxial approximation for the electric field, thus neglecting nonlinear and non-paraxial propagation effects.
According to this simple estimate, in order to increase $\eta_\mrm{THz}$, one should increase the focal beam width $w_0$ while keeping the peak intensity constant, i.e., $E_\mrm{p} \propto w_0^2$. The peak intensity should be chosen such that the gas is fully ionized. Then, because $D_\mrm{p}\propto w_0$ and $L_\mrm{p}\propto w_0^2$, the overall THz pulse energy $E_\mrm{THz}$ is expected to scale with $\propto w_0^6$, and finally $\eta_\mrm{THz} \propto w_0^4 \propto E_\mrm{p}^2$.

Let us finally confront these predictions with rigorous full 3D Maxwell-consistent numerical simulation results. 
We keep all laser and gas parameters as in Fig.~\ref{fig:radprof_pol}, but increase the (vacuum) focal beam width ($w_0=0.8\ldots4.5\,\mu$m) as well as the laser pulse energy ($E_\mrm{p}=0.17\ldots5.4~\mu$J), such that the (vacuum) peak intensity stays constant. 
Figure~\ref{fig:rad_prof}(b) indeed confirms a dramatic enhancement of the laser-to-THz conversion efficiency $\eta_\mrm{THz}$ by more than two orders of magnitude. Our simulation results suggest that, under the right focusing conditions, for only 10-$\mu$J laser energy a conversion efficiency well above $10^{-4}$ can be achieved. 


In our numerical simulations we find, however, a scaling of $\eta_\mrm{THz}$ below the expected $\propto w_0^4$, as clearly visible in Fig.~\ref{fig:rad_prof}(b).
This effect can be attributed to laser defocusing by the free electrons, i.e., a nonlinear propagation effect, which becomes the more pronounced the longer the plasma is. For the highest pulse energy of $5.4~\mu$J, already a 190-$\mu$m-long and 6-$\mu$m-thick plasma is created. The laser defocusing decreases in turn the peak electron density by a factor $\sim3$~(see App.~\ref{app:plasma_profiles}). In order to make a more precise evaluation of the scaling, we renormalize $\eta_\mrm{THz}$ by $(Q/Q^\mrm{wd})^2$, where $Q$ is the final electron charge extracted from the numerical simulation and $Q^{\rm wd} \propto w_0^4$ is the ideal final electron charge without defocusing effects. The resulting scaling $\eta_\mrm{THz}\propto w_0^4\left(Q/Q^\mrm{wd}\right)^2$ is confirmed in Fig.~\ref{fig:rad_prof}(b). 

For the largest pulse energy of 5.4~$\mu$J a 190-$\mu$m-long plasma is created, which qualifies as a line source for THz emission. Line sources emit mostly within a symmetric cone in the laser propagation direction \cite{Kohler:11}. 
In contrast to emission from even longer plasmas~\cite{1367-2630-15-7-075012,PhysRevA.88.063804}, this cone is not hollow, simply because the plasma is too short for the difference of group velocity of the pump in the plasma and phase velocity of THz radiation in the ambient gas to play a significant role.


In summary, we have investigated THz emission from two-color fs-laser-induced microplasmas by means of 3D Maxwell-consistent simulations.
In the strongest focusing case, microplasmas can act as a point source of THz radiation, featuring a toroidal radiation pattern. Because of that, the orientation of the THz emitting current can be determined directly from the radiation profile. This additional information is expected to facilitate the identification of the THz generating mechanism in future experiments. We have shown that the spectral width of the THz emission is broadened up to the maximum plasma frequency in the microplasma, corresponding to the peak free electron density, opening a way for experimental determination of electron densities in microplasmas. Careful theoretical analysis reveals that THz emission from two-color microplasma features an outstanding scalability of the laser-to-THz conversion efficiency. We have developed a simple model which allows to optimize the focusing conditions for a given laser pulse energy. By using the optimized configuration, we have demonstrated that for laser pulse energies below $10~\mu$J, the conversion efficiency can exceed $10^{-4}$. Increasing the conversion efficiency goes hand in hand with a transition to forward radiation from a longer plasma line source. We argue that the excellent scalability of the laser-to-THz conversion efficiency is only possible for transverse radiating THz currents, as they appear for the IC mechanism in the two-color pump configuration. THz generation mechanisms relying on longitudinal radiating currents are detrimental for the upscaling by increasing the plasma length. We believe that our numerical demonstration and analysis of efficient broadband laser-induced THz generation from two-color microplasmas is an important step towards the realization of small-scale THz sources based on laser gas interaction, and will trigger further experimental and theoretical efforts in this direction. 

Numerical simulations were performed using computing resources at
M\'esocentre de Calcul Intensif Aquitaine (MCIA) and Grand {\'E}quipement
National pour le Calcul Intensif (GENCI, Grants No.~2016-056129, No.~2016-057594, and No.~A0020507594). This study was supported by ANR
(Projet ALTESSE). S.S.\ acknowledges support by the Qatar National Research Fund through the National Priorities Research Program (Grant No.\ NPRP 8-246-1-060).

\begin{appendix}
	\section{\label{app:num}Modeling the ionization current mechanism for microplasmas}

The THz generation by two-color laser pulses considered here is driven by the so-called ionization current (IC) mechanism~\cite{Kim}. A comprehensive model based on the fluid equations for electrons describing THz emission has been derived in~\cite{PhysRevE.94.063202}. In this framework, the IC mechanism naturally appears at the lowest order of a multiple-scale expansion. Besides the IC mechanism, this model is also able to treat THz generation driven by ponderomotive forces and others. They appear at higher orders of the multiple-scale expansion and are neglected here. In the following, we briefly summarize the resulting equations governing the IC mechanism.

The electromagnetic fields $\Evec$ and $\Bvec$ are governed by Maxwell's equations
\begin{align}
	\nabla\times\Evec &= - \partial_t \Bvec
	\label{eq:Far_1}\\
	\nabla\times\Bvec &= \frac{1}{c^2}\partial_t \Evec + \mu_0 \Jvec + \mu_0 \Jvec_\mrm{loss}\,\mbox{.}
	\label{eq:Amp_1}
\end{align}
The plasma and electromagnetic fields are coupled via the conductive current density $\Jvec$ governed by
\begin{align}
	\partial_t \Jvec + \nu_\mrm{ei}\Jvec &= \frac{q_\mrm{e}^2}{m_\mrm{e}} n_\mrm{e} \Evec\,\mbox{,}\label{eq:cont}
\end{align}
where electron-ion collisions lead to a damping of the current. The collision frequency is determined by~\cite{Huba2013}
\begin{equation}
	\nu_\mrm{ei}[\mrm{s}^{-1}] = \frac{3.9 \times 10^{-6}\sum\limits_Z Z^2 n_\mrm{ion}^{(Z)}[\mrm{cm}^{-3}]\lambda_\mrm{ei}}{E_{\mrm{elec}}[\mrm{eV}]^{3/2}}  \,\mbox{.}
	\label{eq:nu_ei_NRL_2}
\end{equation}
where $\lambda_\mrm{ei}$ is the Coulomb logarithm. The value $\lambda_\mrm{ei}=3.5$ turned out to match the results obtained by more sophisticated calculations with particle-in-cell (PIC) codes in~\cite{PhysRevE.94.063202}. The densities of $Z$ times charged ions are determined by a set of rate equations
\begin{equation}
\begin{split}
	\partial_t n_\mrm{ion}^{(Z)} & = W^{(Z)} n_\mrm{ion}^{(Z-1)} - W^{(Z+1)} n_\mrm{ion}^{(Z)} \\
	\partial_t n_\mrm{ion}^{(0)} & = - W^{(1)} n_\mrm{ion}^{(0)}
	\label{eq:rate_eq_ion_dens}
\end{split}
\end{equation}
for $Z=1,2,3,\ldots,K$, and the initial neutral density is $n_\mrm{ion}^{(0)}(t=-\infty) = n_\mrm{a}$. The tunnel ionization rate $W^{(Z)}$ in quasi-static approximation creating ions with charge $Z$ is taken from~\cite{Ammosov-1986-Tunnel,PhysRevA.64.013409}. Thus, $W^{(Z)}$ is a function of the modulus of the electric field $\Evec$. The atoms can be at most $K$ times ionized and thus $W^{(K+1)}=0$. The electron density is determined by the ion densities
\begin{equation}
	n_\mrm{e} = \sum_{Z} Z n_\mrm{ion}^{(Z)}\,\mbox{.}
	\label{eq:el_dens_expl}
\end{equation}
The electron energy density $\mathcal{E}=n_\mrm{e}E_{\mrm{elec}}$ is governed by
\begin{equation}
	\partial_t \mathcal{E}=\Jvec\cdot\Evec\,\mbox{.}
\end{equation}
The loss current accounting for ionization losses in the laser field is taken into account by
\begin{align}
	\Jvec_\mrm{loss} = \frac{\Evec}{|\Evec|^2}\sum_Z I_\mrm{p}^{Z} W^{(Z)} n_\mrm{ion}^{(Z-1)}\,\mbox{,} 
\end{align}
where $I_\mrm{p}^{Z}$ is the ionization potential for creation of a $Z$ times charged ion. Even though ionization losses as well as higher order ionization ($Z=2,3,\ldots$) are negligible in the framework of the present study, both are kept for completeness. 

\begin{figure}[b]
	\centering
	\includegraphics[width=0.55\columnwidth]{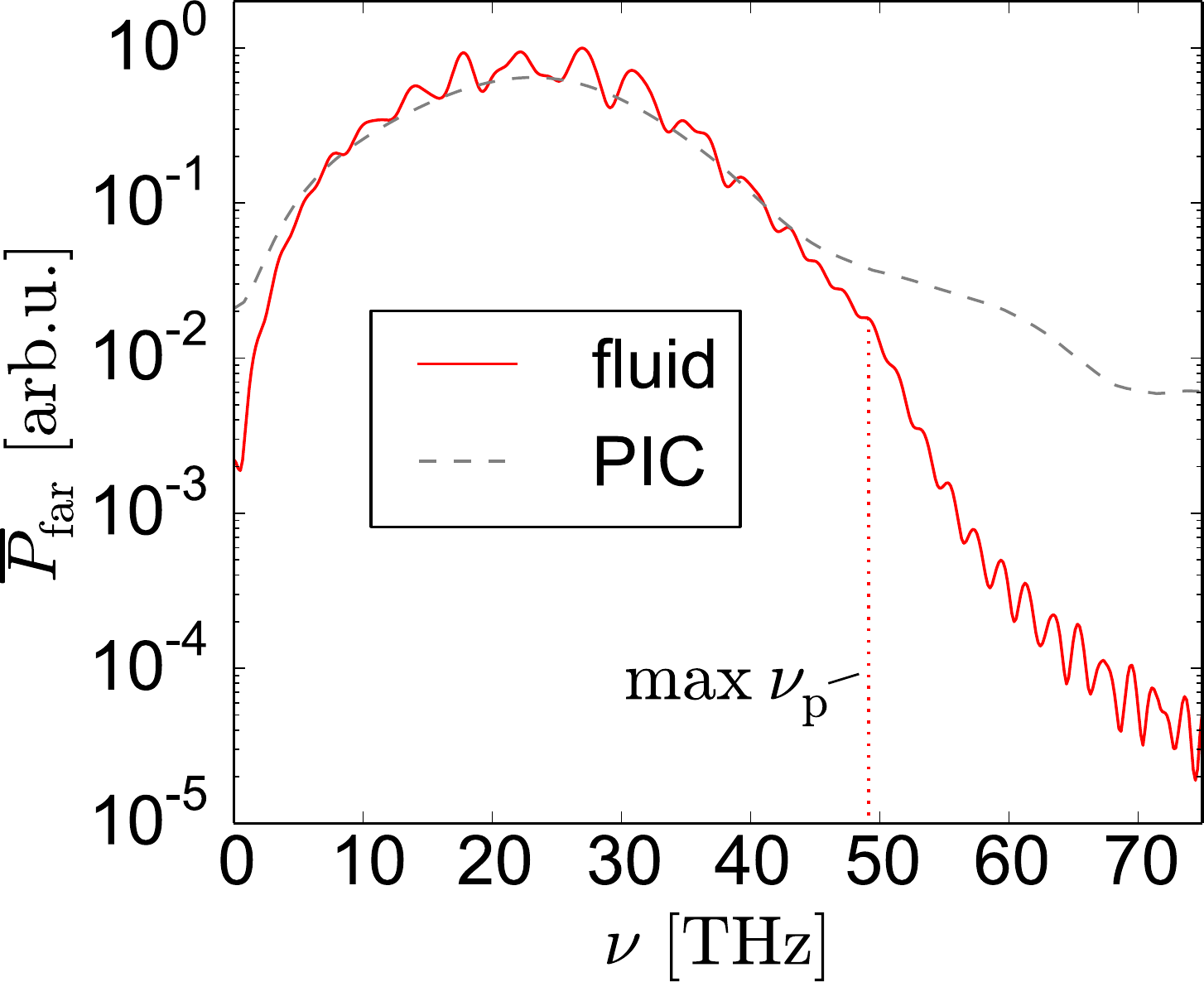}
	\caption{Angularly integrated THz far field spectrum obtained from the fluid model (solid red line) and a PIC simulation (dashed gray line) in 3D.
	The red curve is the same as the one in Fig.~3(b) (iii) of the main article.}
	\label{fig:spec_fluid_PIC}
\end{figure}

The model is implemented in the code ARCTIC, that solves Eqs.~(\ref{eq:Far_1}) and (\ref{eq:Amp_1}) by means of the Yee scheme~\cite{Yee}. We have benchmarked the code ARCTIC by the PIC code OCEAN accounting for full kinetics of the plasma~\cite{PhysRevE.87.043109}. Laser and gas parameters used in this benchmark are given in the caption of Fig.~2 in the main article. The resulting angularly integrated THz far-field spectra are presented in Fig.~\ref{fig:spec_fluid_PIC}. A very good agreement between the fluid based model (solid red line) and the PIC result (dashed gray line) down to the PIC noise level is reported.  

Finally, we comment on computational aspects that turned out to be important for our problem. The fluid simulations with the code ARCTIC have been performed with the spatial resolution of $\delta z = 8$~nm, $\delta x = \delta y = 32$~nm and temporal resolution $\delta t = 25$~as. The resolution in the laser propagation $z$ direction has been selected to be finer than in the transverse directions for the following reason: The use of the Yee scheme for the Maxwell solver implies the strongest numerical dispersion along the $x,y,z$~-~axes \cite{Taflove1995,DSNuter}. When the two-color laser pulse propagates along the $z$~-~axis, the numerical dispersion introduces an artificial phase-shift between the fundamental harmonic (FH) and second harmonic (SH) field. We verified by various simulations in vacuum that this artificial phase-shift is negligible against phase-shifts appearing due to the Gouy-phase for FH and SH fields (see Sec.~\ref{app:phase}) when choosing the resolution that is presented above. The simulation with the PIC code OCEAN has been performed with the spatial resolution of $\delta_x=\delta y = \delta_z = 32$~nm and temporal resolution $\delta t = 106$ as. Because in OCEAN the Maxwell solver is based on the direction splitting (DS) scheme, the numerical dispersion along the $x,y,z$~-~axes is null~\cite{DSNuter} and the resolution in the laser propagation direction is less critical. As a consequence, also the time step could be increased compared to the fluid simulation (Courant-Friedrichs-Lewy condition). The PIC simulation for Fig.~\ref{fig:spec_fluid_PIC} required 58900 CPU hours, while the fluid simulation consumed 4700 CPU hours, which corresponds to a speed-up factor of 12.5. 
Besides the speed-up, in the fluid code ARTIC the fields inside the plasma are significantly less noisy than in PIC simulations (not shown here). For PIC simulations, just the angularly integrated far-field spectra give reliable result in the THz domain due to averaging, while the fields in the plasma are often too noisy to inspect their spatial structure. 

\section{\label{app:plasma_profiles}Plasma profiles and defocusing of the laser pulse}

\begin{figure}[t]
	\centering
	\includegraphics[width=0.85\columnwidth]{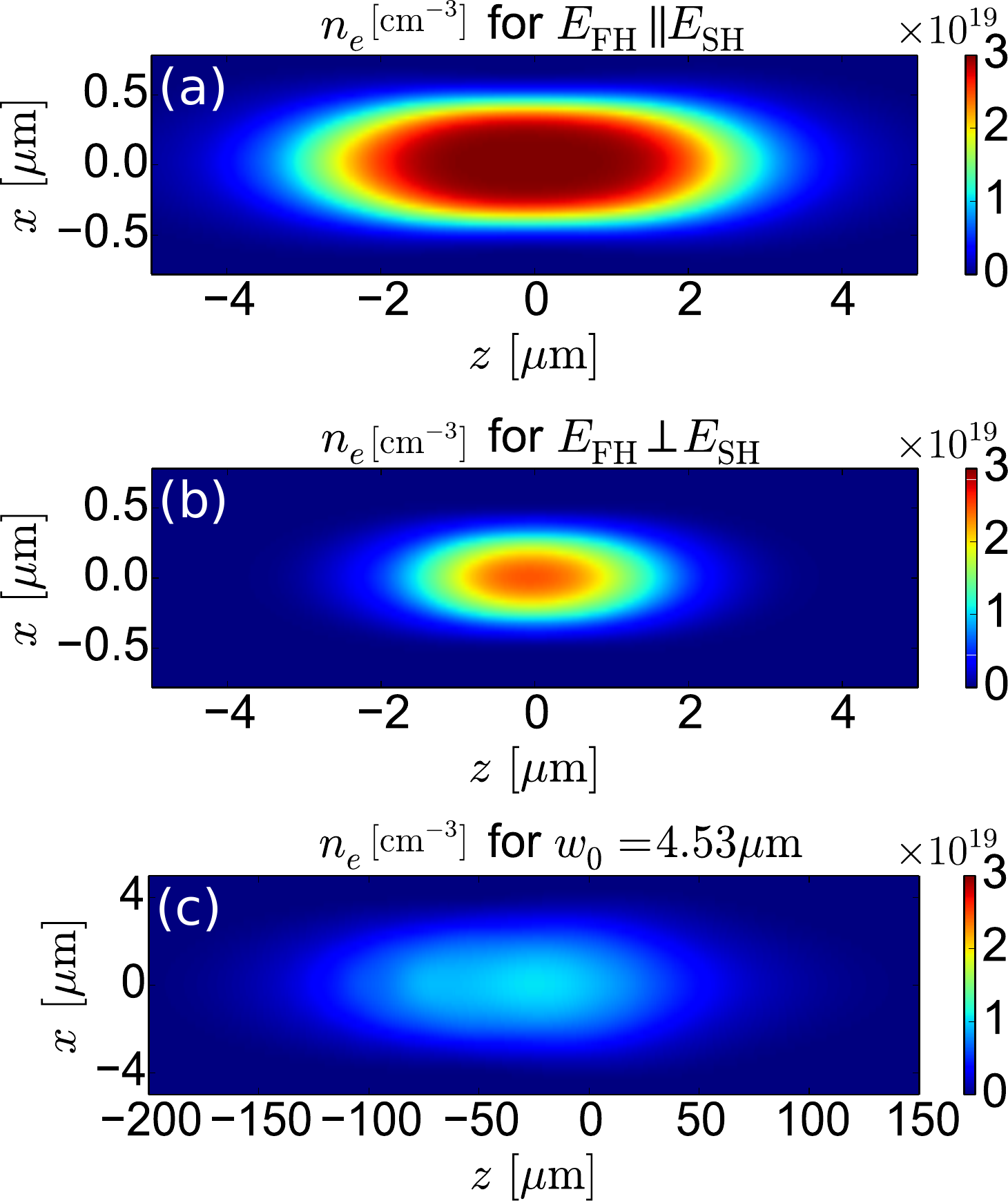}
	\caption{Electron densities after the ionizing laser pulse has passed. The initial argon neutral density is $n_\mrm{a}=3\cdot10^{19}$~cm$^{-3}$ ($p\approx1$~bar). In (a,b) the 0.17-$\mu$J Gaussian laser pulse (20\% energy in the SH, $t_0=50$~fs) is focused down to $w_0=\lambda_\mathrm{FH}=0.8~\mu$m with SH electric field parallel (a) and perpendicular (b) to the FH electric field, respectively. In (c), the focal beam width is increased to $w_0=\lambda_\mathrm{FH}=4.53~\mu$m while keeping the vacuum peak intensity constant and thus increasing the pulse energy to $5.4~\mu$J (SH field is parallel to the FH field).}
	\label{fig:n_e}
\end{figure}

\begin{figure}[t]
	\centering
	\includegraphics[width=0.75\columnwidth]{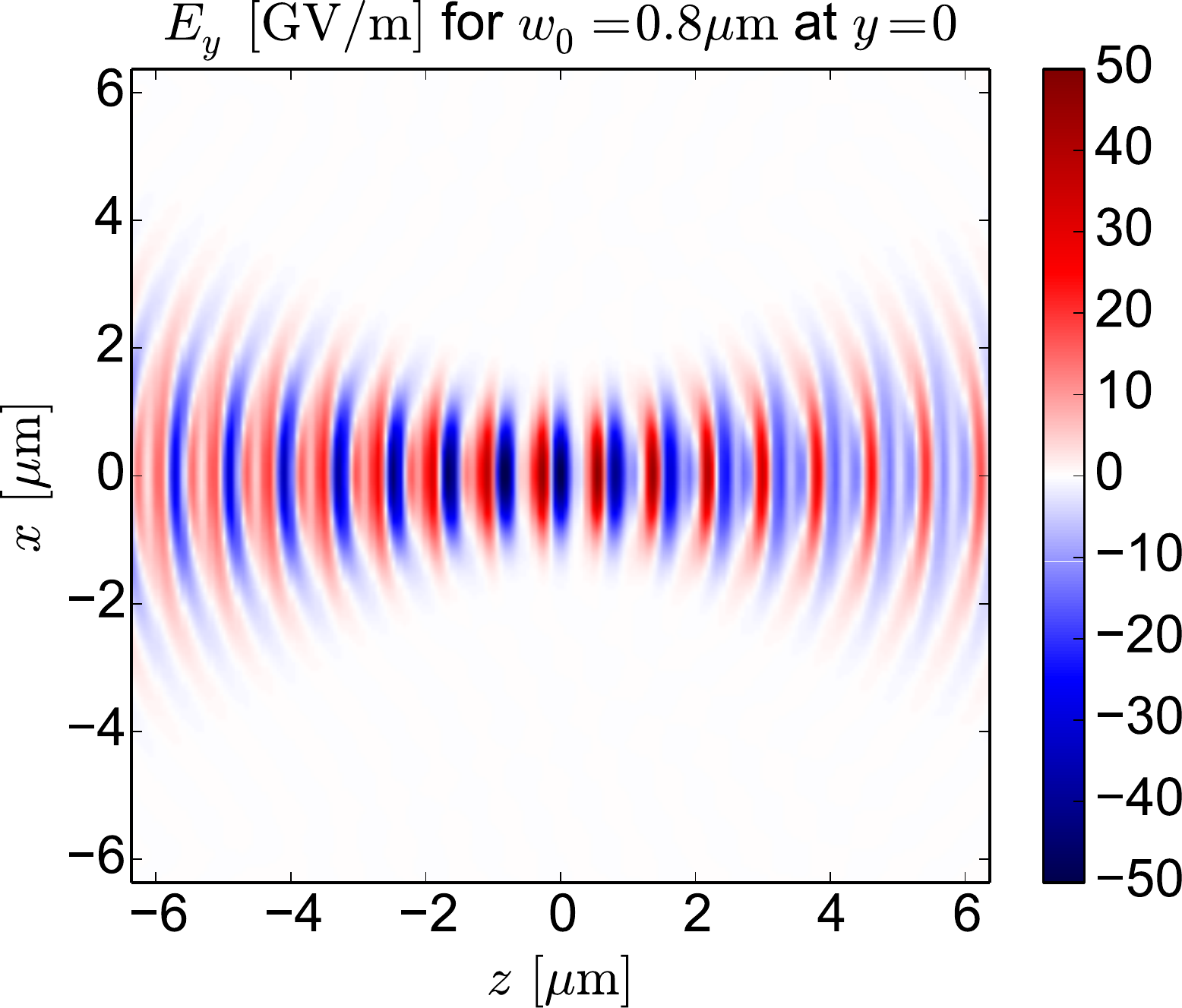}
	\caption{A single snapshots of the electric field $E_x$ for laser and gas parameter same as in Fig.~\ref{fig:n_e}(a).}
	\label{fig:Ey_w0.8}
\end{figure}

\begin{figure*}[t]
	\centering
	\includegraphics[width=0.95\textwidth]{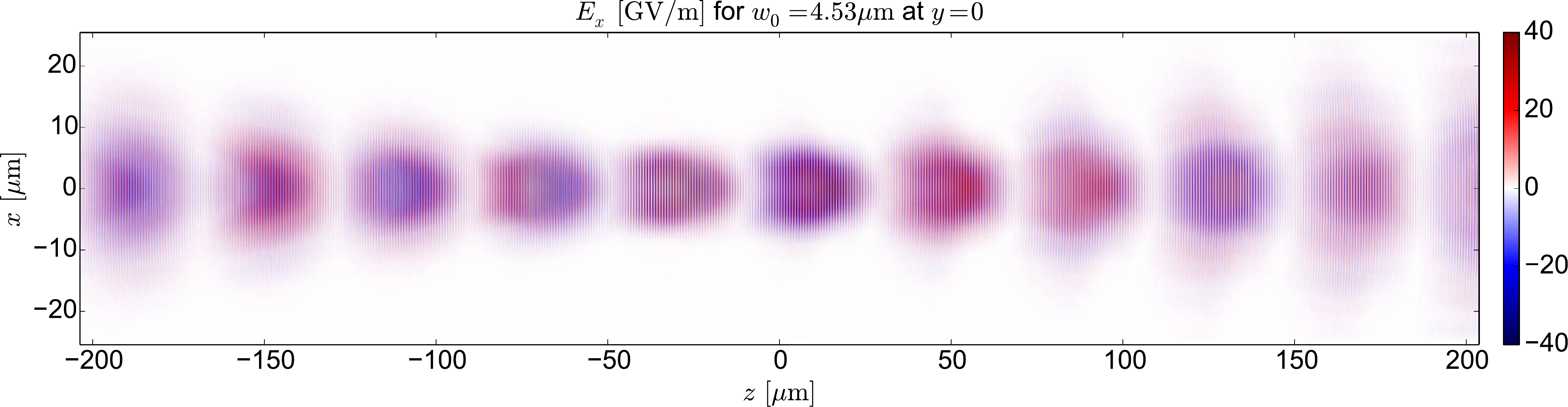}
	\caption{Series of snapshots of the electric field $E_x$ for a laser with $w_0=4.5\,\mu$m and laser pulse energy $E_\mrm{L}=5.4\,\mu$J. The other laser and gas parameter are the same as in Fig.~\ref{fig:n_e}(c).}
	\label{fig:Ey_w4.53}
\end{figure*}

When introducing the strongly focused Gaussian two-color laser pulse of Fig.~1 in the main article ($E_\mrm{p}=0.17~\mu$J, $w_0=0.8~\mu$m, $\evec_\mrm{FH}\parallel\evec_\mrm{SH}$) into argon at ambient pressure, a less that 10-$\mu$m-long and 1-$\mu$m-thick microplasma is created [see Fig.~\ref{fig:n_e}(a)]. The gas is fully singly ionized at $\rvec=0$ in the focal plane, leading to peak electron densities of $n_\mrm{e}=3\times10^{19}$~cm$^{-3}$. Albeit the relatively high electron density, the laser pulse is almost unperturbed by the plasma (see Fig.~\ref{fig:Ey_w0.8}) because of the small interaction length, and propagates almost as in vacuum.

When, in contrast to the configuration in Fig.~\ref{fig:n_e}(a), the SH field is perpendicularly polarized to the FH electric field ($\evec_\mrm{FH}\perp\evec_\mrm{SH}$), the electron density is reduced considerably [see Fig.~\ref{fig:n_e}~(b)]. The reason for this is simply the lower electric field amplitude, because $\sqrt{E_\omega^2+E_{2\omega}^2}\leq E_\omega+E_{2\omega}$. As a consequence, the total electron charge $Q$ is reduced by a factor $\sim 4.2$. Also in this case, the laser pulse is almost unperturbed by the plasma. For more details on the influence of the pump polarization on the THz generation efficiency see Sec.~\ref{app:excit}.

For higher pulse energies and longer plasmas, as exploited in the main paper in order to increase the laser-to-THz efficiency, plasma-induced propagation effects become important. In Fig.~\ref{fig:n_e}(c) the largest plasma we have simulated is shown, created by a $5.4$-$\mu$J two-color pulse ($w_0=4.53\,\mu$m). Compared to Fig.~\ref{fig:n_e}(a), the peak electron density is reduced by a factor $\sim 3$. The reason is plasma defocussing, which becomes a key player now. In Fig.~\ref{fig:Ey_w4.53} a series of snapshots of the laser pulse is presented and allows to track the defocussing process. The laser pulse is focused several tens of micrometers before the vacuum focus at $z=0$. The peak electric field is reduced from $\sim 50$~GV/m to $\sim 40$~GV/m. 

\section{\label{app:phase}Phase angle $\phi$ between FH and SH}

The excitation of the THz current from the ionization current (IC) mechanism is driven by the term $n_\mrm{e}\Evec$ in equation Eq.~(\ref{eq:cont}). For microplasmas, $n_\mrm{e}\Evec_\mrm{L}$ can be considered as the exciting source term, where $\Evec_\mrm{L}$ is the electric laser field in vacuum~\cite{PhysRevE.94.063202}. The laser field $\Evec_\mrm{L}$ is the sum of the FH field and the SH field, and it is well established that the strongest excitation at THz frequencies takes place for the relative angle $\phi=\pi/2$ between FH and SH~\cite{Kim}. This is why we choose $\phi=\pi/2$ at the vacuum focus in all our simulations [see also Eq.~(2) of the main article].

\begin{figure}[b]
	\centering
	\includegraphics[width=.8\columnwidth]{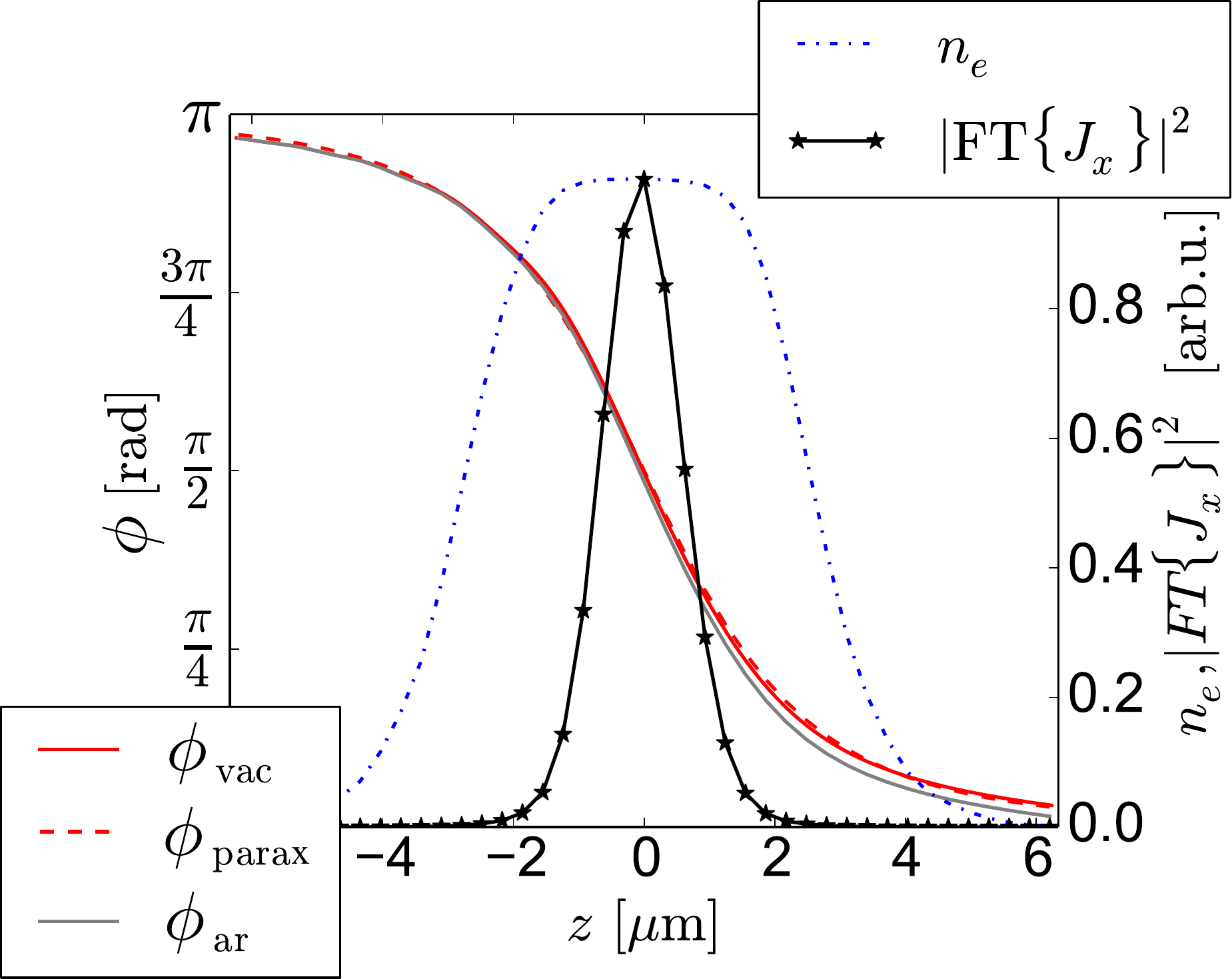}
	\caption{Relative phase $\phi(z)$ between the FH and SH field for laser and gas parameter same as in Fig.~\ref{fig:n_e}(a) on the optical axis ($x=y=0$) as well as the electron density $n_\mrm{e}(z)$ (after the pulse has passed) and the on-axis peak power spectum of the THz current ($<60$~THz). The electron density $n_\mrm{e}$ and the low frequency current source are normalized to one.}
	\label{fig:phase}
\end{figure}

However, the FH and SH fields are propagating differently, even in vacuum. 
This makes in particular the relative phase angle $\phi$ between the two harmonics $z$-dependent
\begin{equation}
	\phi(z) = \psi(z,2\omega_\mrm{L})-2\psi(z,\omega_\mrm{L}),
\end{equation}
where $\psi$ is the $z$-dependent phase at the corresponding frequency. The variation of $\phi$ along $z$ implies a strong variation of the excitation efficiency of the THz current along $z$ (see Sec.~\ref{app:excit} and the following example). 

The variation of $\psi$ is determined well in the paraxial approximation by the Gouy phase such that
\begin{equation}
	\psi_\mrm{parax}(z,\omega) = \psi_\mrm{parax}(z=0,\omega) + \arctan\left(\frac{2cz}{w_0^2 \omega}\right)\,\mbox{,}
\end{equation}
where with our definition of the laser in the main article $\psi_\mrm{parax}(z=0,\omega_\mrm{L}) = 0$ and $\psi_\mrm{parax}(z=0,2\omega_\mrm{L}) = \phi(z=0)$. Let us consider the strongest focusing case as presented in Fig.~(\ref{fig:Ey_w0.8}). The dark red dashed line in Fig.~\ref{fig:phase} is presenting the corresponding $\phi_\mrm{parax}$, which almost coincides with both vacuum Maxwell and argon gas Maxwell simulation results. At focus, the optimum value of $\phi=\pi/2$ is retrieved. However, the phase angle $\phi$ changes from $\pi$ to $0$ over a couple of $\mu$m propagation range.  
Because only $\phi=\pi/2$ leads to an optimum excitation of the THz current, for $z\neq 0$ the excitation is significantly weaker. As a result, the low-frequency current source presented in Fig.~\ref{fig:phase} (solid black line with star markers) is significantly shorter than the plasma, indicated by its electron density (dash-dotted blue line).

\begin{figure*}
  \includegraphics[width=1.\textwidth]{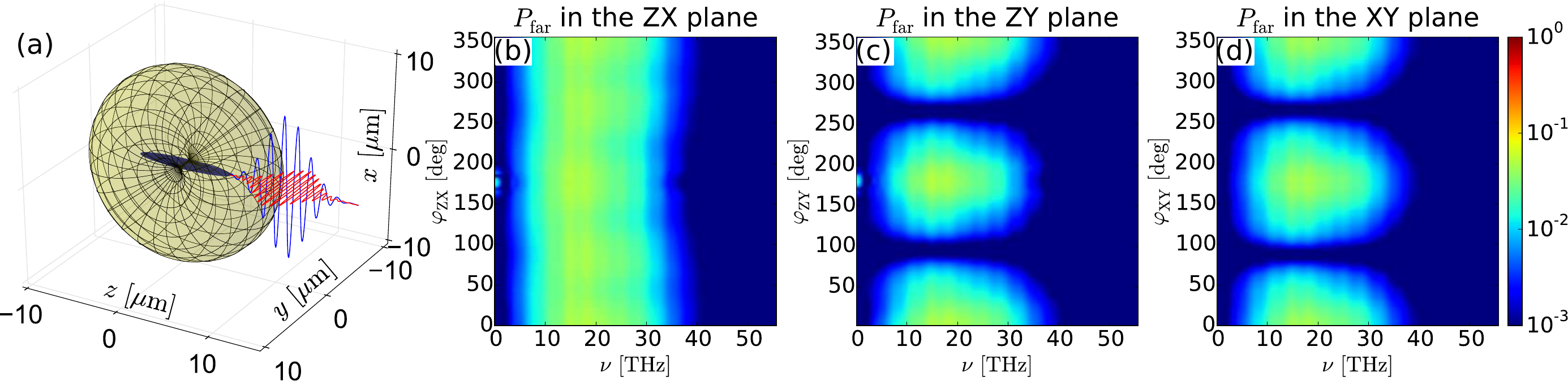}
  \caption{The same two-color 0.17-$\mu$J Gaussian laser pulse as in Fig.~2 of the main article, but FH and SH electric fields are  perpendicularly polarized ($\evec_\mrm{FH} = \evec_x$, $\evec_\mrm{SH} = \evec_y$). The radiation pattern for one THz frequency (assuming a point source at $\rvec=0$) is sketched in (a) (green surface) as well as the plasma (blue surface). In (b-d) the simulated THz radiation profiles in terms of the far field power spectra $P_{\rm far}$ in the ZX ($y=0$), ZY ($x=0$), and XY ($z=0$) planes are presented. Color scales are comparable with those in Fig.~2 of the main article, confirming a lower THz energy yield for perpendicular pump polarization.}
  \label{fig:radprof_pol_2}
\end{figure*}

\section{\label{app:excit}Polarization of FH and SH fields}

\begin{figure}[b]
	\centering
	\includegraphics[width=1.\columnwidth]{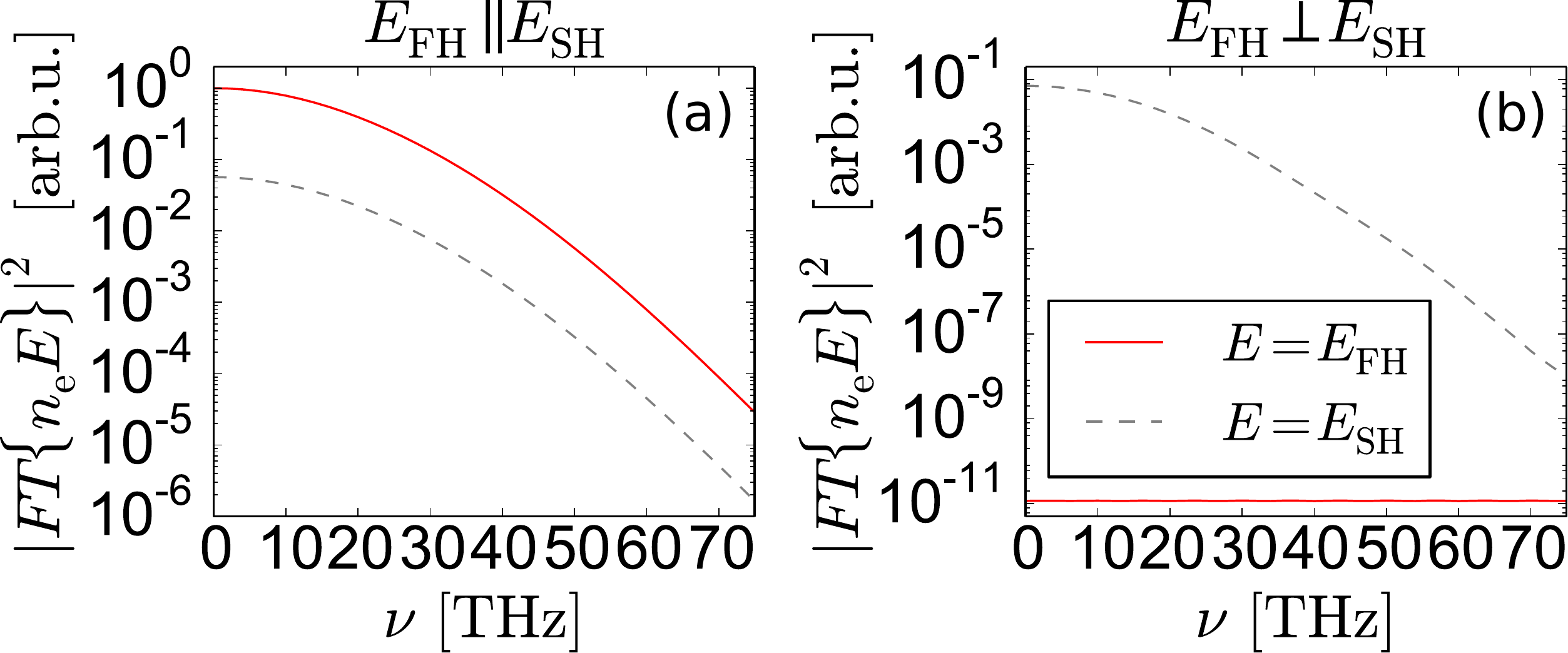}
	\includegraphics[width=1.\columnwidth]{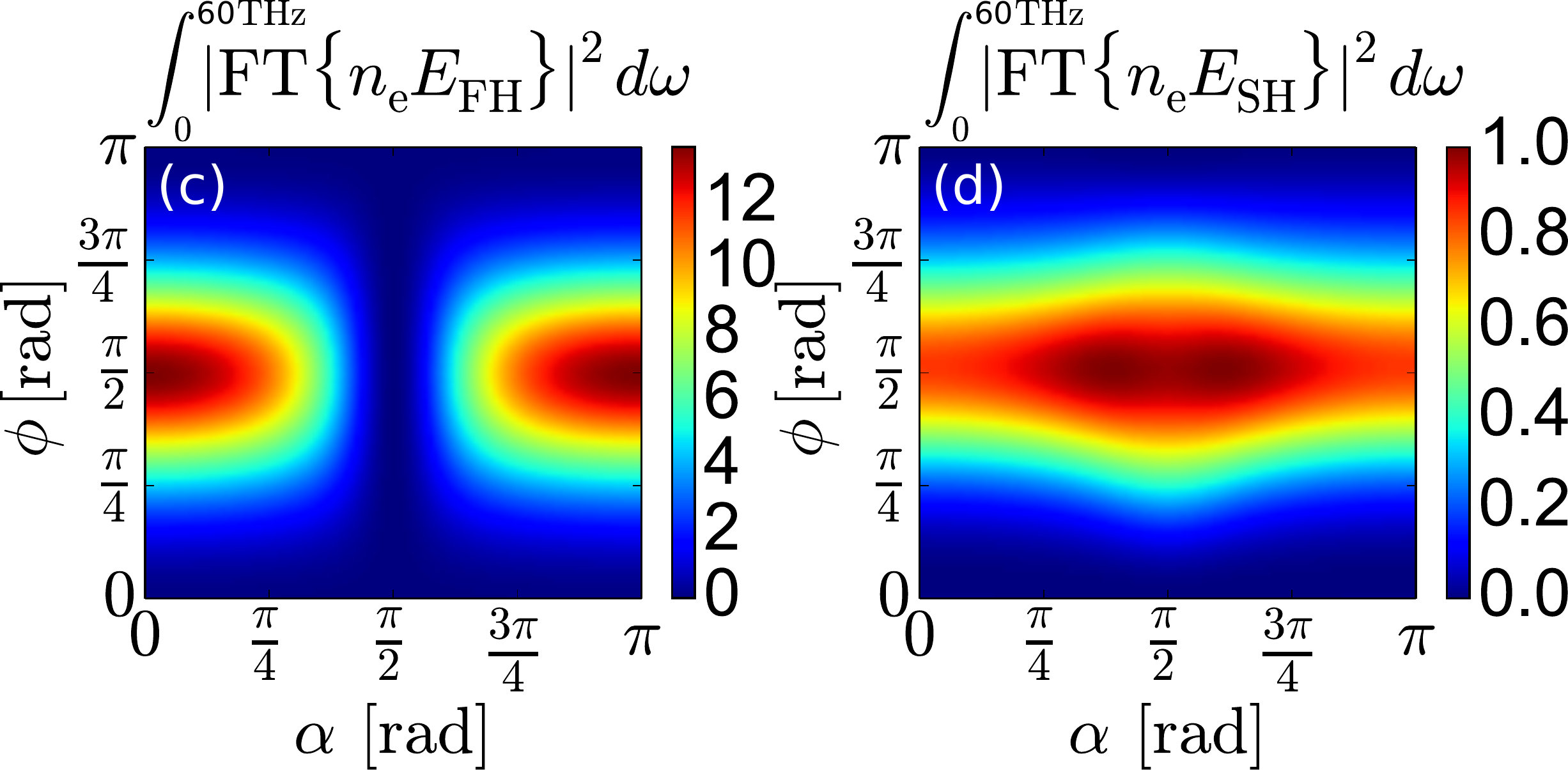}	
	\caption{Power spectrum of the excitation source $n_\mrm{e}\Evec$ at focus where $\Evec$ is the FH part of the laser field (dark solid red line) or SH part (light gray dashed line) for $\phi=\pi/2$, with laser and gas parameter as in Fig.~\ref{fig:n_e}(a); (a) $\alpha=0$ (SH field parallel to the FH field); (b) $\alpha=\phi=\pi/2$ (SH field perpendicular to the FH field). Frequency integrated excitation source driven by the FH (c) and SH (d) field as a function of the angle $\alpha$ between FH and SH field vectors as well as $\phi$.}
	\label{fig:ex}
\end{figure}

\begin{figure}[b]
     \centering
\includegraphics[width=0.99\columnwidth]{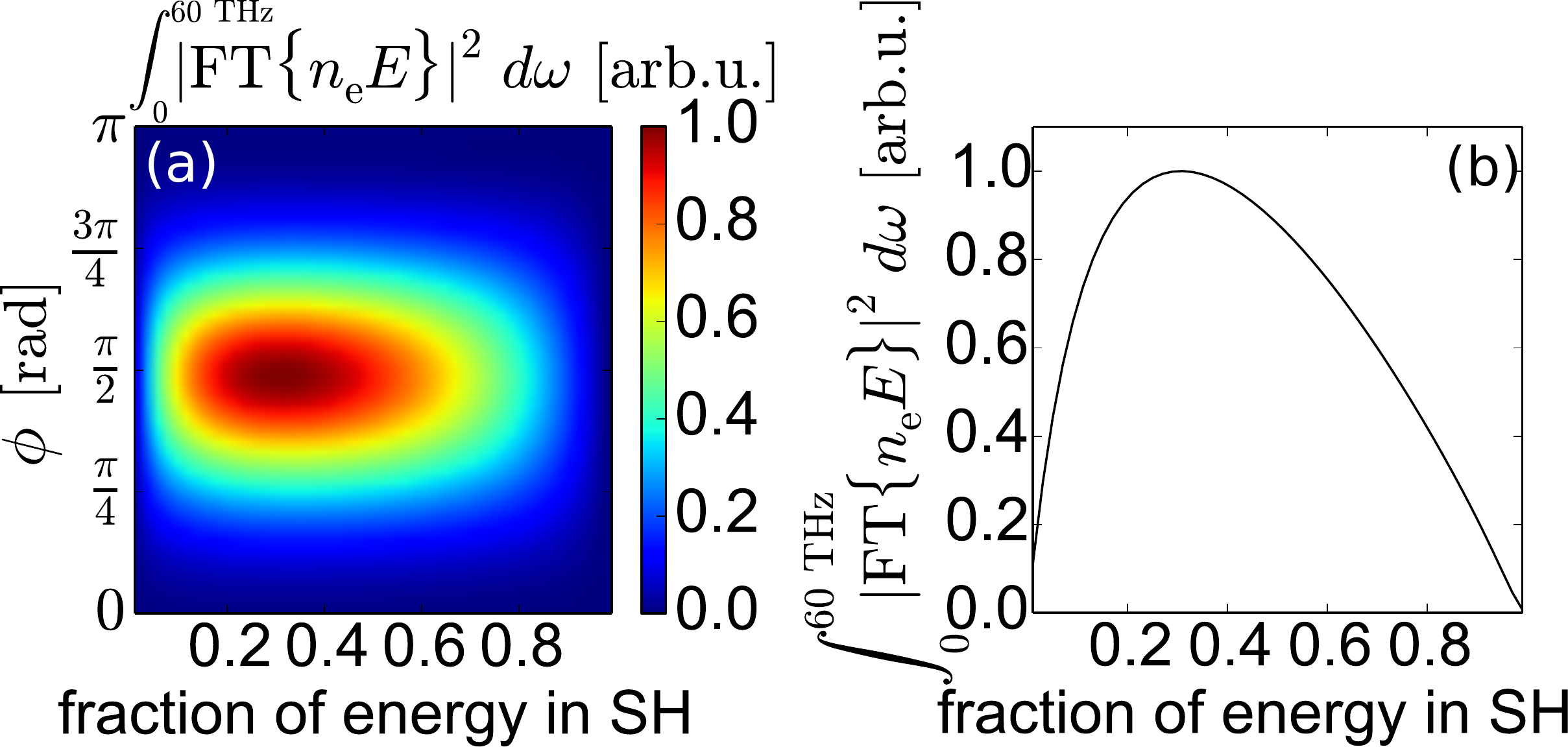}
     \caption{Frequency integrated excitation source $n_\mrm{e}\Evec$ 
for laser and gas parameter as in Fig.~\ref{fig:n_e}(a). In (a) the 
dependence on $\phi$ and the fraction of energy in the SH pulse component 
is presented. In (b) a line-out of (a) for the optimal angle 
$\phi=\pi/2$ is shown: The dependency on the SH pulse energy is not so critical. 
Already for few percent of the total energy in the SH field, the excitation source is reasonably strong.}
     \label{fig:ex_amp}
\end{figure}

The polarization of THz radiation depends on the polarization of the two-color pump laser fields. When the FH and SH electric fields are parallel to each other ($\evec_\mrm{FH} \parallel \evec_\mrm{SH}$, like in all examples given in the main article), the THz emission is linearly polarized in the same direction. Turning the SH polarization by 90 degrees with respect to the FH polarization makes the THz polarization follow the SH one, as shown in Fig.~\ref{fig:radprof_pol_2}.  


The reason for this is explained in the following, where the THz source term $n_\mrm{e}\Evec$ is analyzed componentwise. For the parallel case ($\alpha=0$), the low-frequency power spectrum of $n_\mrm{e}\Evec_\mrm{FH}$ (red solid line) and $n_\mrm{e}\Evec_\mrm{SH}$ (gray dashed line) is presented in Fig.~\ref{fig:ex}(a). Here, the excitation is driven by the FH and the SH electric fields. However, the excitation by the FH electric field dominates. For the perpendicular situation ($\alpha=\pi/2$) shown in Fig.~\ref{fig:ex}(b), solely the SH electric field is responsible for the excitation. The reason is the time-dependence of the electron density $n_\mrm{e}$ that does not provide a FH frequency component in this case (not shown here). Thus, the product of $n_\mrm{e}$ with $\Evec_\mrm{FH}$ does not lead to a low-frequency component of the source. Since the excitation is driven by the SH electric field, the THz-emitting current has the polarization direction of the SH electric field. As detailed in Fig.~\ref{fig:ex}(c-d), the optimum configuration for THz production is $\phi = \pi/2$ and $\alpha=0$.

\section{\label{app:amplitude}Amplitude ratio of FH and SH}

Last but not least we comment on the impact of the FH vs.\ SH amplitude resp.\ energy ratio. In recent experiments, various fractions of the total pulse energy were converted to the SH frequency. In focused geometries 1~\% up to 12~\% have been demonstrated \cite{PhysRevLett.105.053903} and even more than 20~\% are in principle possible when focusing the FH and SH beam after its generation in collimated geometry. Our analysis of the strength of THz source term $n_\mrm{e}\Evec$ depending on the energy ration is presented in Fig.~\ref{fig:ex_amp}. Throughout this work, we assume that 20~\% of the total pulse energy is in the SH field, thus we are very close to the optimum case.

\section{\label{app:cyl}Simple model for THz radiation from a plasma}

Simple considerations on the emission from a radiating current structure can explain various tendencies that we observe in our simulations. 
To this end, we do not consider the origin of the THz radiating current, but just assume that such current exists in a volume with length $L_\mrm{p}$ and width $D_\mrm{p}$.
As pointed out in Sec.~\ref{app:phase}, the radiating current structure is not necessarily as large as the plasma. However, as for the plasma profile itself, we can assume that the scaling of $L_\mrm{p}$ and $D_\mrm{p}$ with the laser pulse parameters is the same as long as the laser pulse is not deformed too much, i.e., we are dealing with a microplasma.

When keeping the peak intensity at focus constant while increasing the focal beam width $w_0$ and laser pulse energy $E_\mrm{p}$ simultaneously, the following assumptions (obtained from quasi-monochromatic paraxial approximation) on the scalings are reasonable~\cite{PhysRevE.94.063202}:
\begin{align}
	D_\mrm{p}  & \propto w_0\,\mbox{,} & L_\mrm{p} & \propto w_0^2\,\mbox{.} \label{eq:Dp_Lp_scaling}
\end{align}
In order to get the scaling of the emitted THz pulse energy $E_\mrm{THz}$ with $L_\mrm{p}$ and $D_\mrm{p}$, the following model is proposed: The radiating current structure is assumed to move invariantly with the vacuum speed of light $c$ (or, more precisely, with the group velocity of the pump pulse) and thus
\begin{equation}
	\Jvec(\rvec_\perp,z,t)=\Jvec_0(t-z/c)
	\label{eq:J_assumption}
\end{equation} 
in a cylinder with diameter $D_\mrm{p}$ and length $L_\mrm{p}$. Outside this cylinder, which is centered around $z=0$, we assume vaccum and thus zero current density. In temporal Fourier space, this assumption in Eq.~(\ref{eq:J_assumption}) translates into
\begin{equation}
	\hat{\Jvec}(\rvec_\perp,z,\omega)=\hat{\Jvec}_0(\omega)e^{\rmi\frac{\omega}{c}z}\,\mbox{.}
	\label{eq:J_assumption_omega}
\end{equation}

In the far-field, the radiated magnetic field follows~\cite{PhysRevE.94.063202}
\begin{equation}
	\hat{\Bvec}_\mrm{far}(\rvec,\omega) \approx -\rmi \mu_0 \frac{\omega}{c} \frac{\rvec}{|\rvec|} \times \! \int\limits_{V}
	\!\hat{\Jvec}(\rvec',\omega) \, \frac{e^{\rmi \frac{\omega}{c} |\rvec-\rvec'|}}{4\pi |\rvec-\rvec'|} \,d^3\rvec'\label{eq:B_far_app}\,\mbox{,}
\end{equation}
where $\rvec=(x,y,z)^\mrm{T}$ is the position of the detector and $V$ is the plasma volume. The far-field power spectrum is given by
\begin{equation}
	P_\mrm{far}(\rvec,\omega)
			   =\frac{c}{\mu_0}\left|\hat{\Bvec}_\mrm{far}(\rvec,\omega)\right|^2\,\mbox{.}
	\label{eq:Pfar:def}
\end{equation}
The goal is now to simplify Eq.~(\ref{eq:Pfar:def}) using Eq.~(\ref{eq:B_far_app}) for currents that fulfill Eq.~(\ref{eq:J_assumption_omega}), i.e.,
\begin{equation}
	P_\mrm{far}
			   =\frac{\mu_0}{c}\omega^2\left|\frac{\rvec}{|\rvec|}\times \! \int\limits_{V}
			   	\!\hat{\Jvec}_0(\omega) \, \frac{e^{\rmi \frac{\omega}{c} \left(|\rvec-\rvec'|+z'\right)}}{4\pi |\rvec-\rvec'|}\,d^3\rvec'\right|^2\,\mbox{.}
	\label{eq:Pfar}
\end{equation}
In order to perform the integration over $\rvec^\prime$ in Eq.~(\ref{eq:Pfar}), cylindrical coordinates $(r_\perp^\prime,\varphi^\prime,z^\prime)$ are introduced as
\begin{align}
	x^\prime  & = r_\perp^\prime \cos{\varphi^\prime} & 
	y^\prime  & = r_\perp^\prime \sin{\varphi^\prime}\,\mbox{.} 
\end{align}
Moreover, we note that the distance $|\rvec - \rvec^\prime|$ expands in the far-field by means of Taylor expansion as
\begin{equation}
	|\rvec - \rvec^\prime| \approx |\rvec|-r^\prime_\perp\frac{x\cos{\varphi^\prime}+y\sin{\varphi^\prime}}{|\rvec|} - z^\prime\frac{z}{|\rvec|}\,\mbox{.}
\end{equation}
Using this expansion in the exponent in Eq.~(\ref{eq:Pfar}) and $|\rvec-\rvec'|\approx|\rvec|$ we obtain
\begin{equation}
\begin{split}
	P&_\mrm{far} \approx \underbrace{\left|\frac{\rvec\times\hat{\Jvec}_0(\omega)}{|\rvec|}\right|^2}_{=:F_\mrm{\Jvec_0}} \underbrace{\left|\,\int\limits_{-\frac{L_\mrm{p}}{2}}^{\frac{L_\mrm{p}}{2}}e^{\rmi\frac{\omega}{c}z'\left(1-\frac{z}{|\rvec|}\right)}\,dz^\prime\right|^2}_{=:F_{L_\mrm{p}}} \\
	& \times \underbrace{\frac{\mu_0\omega^2}{16\pi^2c|\rvec|^2} \left|\int\limits_{A} e^{-\rmi\frac{\omega}{c}r^\prime_\perp
	\frac{x\cos{\varphi^\prime}+y\sin{\varphi^\prime}}{|\rvec|}} r^\prime_\perp \, dr_\perp^\prime\,d\varphi^\prime\right|^2}_{=:F_{D_\mrm{p}}}\,\mbox{,}
	\label{eq:Pfar_1}
\end{split}
\end{equation}
where $A$ is the transverse surface of the plasma. 

It is very interesting to note that for our model-current Eq.~(\ref{eq:J_assumption_omega}), the far-field power spectrum separates into three terms that have their own simple role. In order to discuss their dependencies properly, we now switch to standard (ISO) spherical coordinates $(r,\theta,\phi)$ according to
\begin{equation}
	x = r\sin(\theta)\cos(\phi)\quad
	y = r\sin(\theta)\sin(\phi)\quad
	z = r\cos(\theta)\,\mbox{.}
\end{equation}
\begin{itemize}
\item The term $F_{\Jvec_0}(\theta,\phi,\omega)$ is mainly characterized by the orientation of the current. 
\item The term $F_{L_\mrm{p}}(\theta,\omega)$ is solely determined by the length of the plasma, and can be evaluated as
\begin{equation}
F_{L_\mrm{p}}(\theta,\omega) = \frac{4c^2\sin^2\!\left[\frac{\omega L_\mrm{p}}{2c} (1-\cos\theta) \right]}{\omega^2(1-\cos\theta)^2 }\,\mbox{.}\label{eq:F_Lp}
\end{equation}
\item The term $F_{D_\mrm{p}}(r,\omega)$ is dependent on the transverse profile of the current structure. 
For thin plasmas, i.e. $D_\mrm{p}/\lambda_\mrm{THz}\ll1$ where $\lambda_\mrm{THz}=c/\nu_\mrm{THz}$, it reads 
\begin{equation}
F_{D_\mrm{p}}(r,\omega) \approx \frac{\mu_0\omega^2 D_\mrm{p} ^4}{256cr^2}\,\mbox{.}
\end{equation}
Thus, the THz pulse energy scales as $E_\mrm{THz}\propto D_\mrm{p}^4$.
\end{itemize}

Next, the angularly integrated power spectrum $\overbar{P}_\mrm{far}$ is computed from Eq.~(\ref{eq:Pfar_1}) for thin plasmas and transverse resp.\ longitudinal currents:
\begin{equation}
\begin{split}
	\overbar{P}_\mrm{far} & = \iint\limits_{\Omega} P_\mrm{far}\,r^2 \sin\theta \, d\theta \, d\phi = \frac{\pi\mu_0c D_\mrm{p}^4}{64}|\Jvec_0(\omega)|^2 \\
	&\quad \times\int\limits_0^{\pi} \frac{\sin^2\!\left[\frac{\omega L_\mrm{p}}{2c} (1-\cos\theta) \right]}{(1-\cos\theta)^2 } G_{\Jvec_0}(\theta)
\, \sin\theta \, d\theta\,\mbox{,}
\end{split}
\end{equation}
where
\begin{equation}
G_{\Jvec_0}(\theta) = \begin{cases}
	     1 + \cos^2{\theta} & \text{for $\Jvec_0 \parallel \evec_x$} \\
	     2\sin^2{\theta} & \text{for $\Jvec_0 \parallel \evec_z$}
	   \end{cases} \,\mbox{.}
\end{equation}
For $L_\mrm{p}\ll\lambda_\mrm{THz}$, both cases give by means of Taylor expansion the scaling
$E_\mrm{THz}\propto L_\mrm{p}^2$.
For $L_\mrm{p}\gg\lambda_\mrm{THz}$, numerical evaluation gives 
$E_\mrm{THz}\propto L_\mrm{p}$ for $\Jvec_0 \parallel \evec_x$,
and a sub-linear behavior for $\Jvec_0 \parallel \evec_z$.

\section{\label{app:scale}Scaling laws}

In the following, the scalability of the laser-to-THz conversion efficiency $\eta_\mrm{THz}$ in microplasmas is investigated based on the model developped in Sec.~\ref{app:cyl}. As explained in the main article, we expect optimum conditions for THz generation when the laser pulse energy is increased simultaneously with $E_\mrm{p}\propto w_0^2$, thus keeping the vacuum peak intensity at focus constant. 
Let us assume for a moment that the ionizing laser pulse is not distorted by the produced plasma, i.e., Eq.~(\ref{eq:Dp_Lp_scaling}) holds. Then, the results obtained in Sec.~\ref{app:cyl} suggest that the THz pulse energy scales as $E_\mrm{THz}\propto w_0^6$ for longer plasmas ($L_\mrm{p}\gg\lambda_\mrm{THz}$) and therefore $\eta_\mrm{THz} = E_\mrm{THz}/E_\mrm{p} \propto w_0^4$. One would also expect a scaling of the final electron charge as $Q^{\rm wd}\propto L_\mrm{p}D_\mrm{p}^2\propto w_0^4$.

However, longer plasmas lead to laser defocusing as discussed in Sec.~\ref{app:plasma_profiles}, which reduces the electron charge and thus the number of THz emitters. The simulated values of the electron charge $Q$ versus $w_0$ and resulting exponents $k_Q$ obtained from nearest neighbours fitting~\footnote{$k_Q = \log\!\left(Q^+/Q^-\right)/\log\!\left(w_0^+/w_0^-\right)$, where superscripts $+$ and $-$ refer to left and right neighbour, respectively.} are presented in Tab.~\ref{tab}. 
Between $w_0=0.8\,\mu$m and $w_0=1.13\,\mu$m, we find the expected exponent $k_Q \approx 4$. 
The corresponding exponent for the laser-to-THz conversion efficiency $k_\eta$ is even slightly higher than 4, because we are at the limit to short plasma scaling with $E_\mrm{THz}\propto L_\mrm{p}^2$. 
For larger $w_0$, the exponent $k_Q$ decreases significantly. The exponent $k_\eta$ decreases even stronger than $k_Q$, and the question is whether this decrease can be explained by the plasma defocusing only, or some additional parasitic effect is at play.

Even in simulations with strong defocusing, we observe that $L_\mrm{p}\propto w_0^2$ and $D_\mrm{p}\propto w_0$ for the normalized electron density profile, and only the number of created electrons, i.e., the density $n_e$, is decreasing. 
Because the term $n_\mrm{e}\Evec$ driving the current $\Jvec$ is proportional to $n_e$, we can expect that the conversion efficiency $\eta_\mrm{THz}$ gets simply reduced by a factor $\propto n_e^2 \propto (Q/Q^{\rm wd})^2$.
In order to verify this hypothesis, we introduce the rescaled conversion efficiency $\tilde{\eta}_\mrm{THz} = \eta_\mrm{THz}/(Q/Q^{\rm wd})^2$.
The corresponding exponent $k_{\tilde{\eta}}$ is presented in the last line of Tab.~{\color{red}I}, and seems to support the scaling law
\begin{equation}
	\eta_\mrm{THz} \propto w_0^4\left(\frac{Q}{Q^{\rm wd}}\right)^2\,\mbox{.}
\end{equation} 
Thus, the sub-optimal scaling of $\eta_\mrm{THz}$ in the microplasmas considered here (up to $L_\mrm{p}=190~\mu$m) can be attributed solely to plasma defocusing.

\renewcommand{\arraystretch}{1.3}
\begin{table}
	\begin{tabular}{ | m{2cm} ? c  c | c  c | c  c | c  c | c  c | c  c |} \hline
		$w_0\,[\mu\mathrm{m}]$ & \multicolumn{2}{c|}{0.8}& \multicolumn{2}{c|}{1.13}& \multicolumn{2}{c|}{1.6}& \multicolumn{2}{c|}{2.26}& \multicolumn{2}{c|}{3.2}& \multicolumn{2}{c|}{4.53} \\ \hline
		
		$Q\,[\mathrm{nC}]$ & \multicolumn{2}{c|}{0.014}& \multicolumn{2}{c|}{0.057}& \multicolumn{2}{c|}{0.21}& \multicolumn{2}{c|}{0.67}& \multicolumn{2}{c|}{2.05}& \multicolumn{2}{c|}{5.42} \\ \hline
		
		$\eta_\mrm{THz}\,[10^{-5}]$ & \multicolumn{2}{c|}{0.074}& \multicolumn{2}{c|}{0.338}& \multicolumn{2}{c|}{1.25}& \multicolumn{2}{c|}{3.65}& \multicolumn{2}{c|}{8.26}& \multicolumn{2}{c|}{14.4} \\ \hline
		
		$k_Q$ & \,\,\,\,\,\,\, & \multicolumn{2}{c|}{\,4.05\,\,}& \multicolumn{2}{c|}{3.69\,}& \multicolumn{2}{c|}{3.43\,}& \multicolumn{2}{c|}{3.20}& \multicolumn{2}{c}{\,2.80}&\\ \hline
		
		$k_\eta$ & & \multicolumn{2}{c|}{4.39}& \multicolumn{2}{c|}{3.75}& \multicolumn{2}{c|}{3.11}& \multicolumn{2}{c|}{2.35}& \multicolumn{2}{c}{1.60}&\\ \hline
		
		$k_{\tilde{\eta}}$ & & \multicolumn{2}{c|}{4.29}& \multicolumn{2}{c|}{4.36}& \multicolumn{2}{c|}{4.24}& \multicolumn{2}{c|}{3.94}& \multicolumn{2}{c}{4.00}&\\ \hline
	\end{tabular}
	\caption{Simulations results varying the focal beam width $w_0$ for a constant vacuum peak intensity and thus increasing the laser pulse energy. For $w_0=0.8\,\mu$m, the laser and gas parameter are the same as in Fig~\ref{fig:spec_fluid_PIC}.}
	\label{tab}
\end{table}

\end{appendix}

\bibliography{mybibfile}

\end{document}